\documentclass[aps,twocolumn,amsmath,amssymb]{revtex4-2}
\usepackage[dvipdfmx]{graphicx}
\usepackage{dcolumn}
\usepackage{bm}
\usepackage{braket}
\usepackage{hyperref}
\hypersetup{
colorlinks = true, 
urlcolor = blue, 
linkcolor = blue, 
citecolor = blue
}

\begin{document}

\title{
Floquet engineering of effective pairing interactions in a doped band insulator
}
\author{
Yugo Takahashi, 
Hideo Miyamoto, 
Kazuhiko Kuroki, 
and Tatsuya Kaneko
}
\affiliation{
Department of Physics, Osaka University, Toyonaka, Osaka 560-0043, Japan
}
\date{\today}
\begin{abstract}
We investigate the pairing state in a doped band insulator under a periodic driving field. 
We employ a correlated fermionic model on a honeycomb lattice, in which pairing glue is obtainable via repulsive interactions, and derive an effective model under circularly polarized light. 
We demonstrate that the effective pairing interaction for doped fermions obtained with the second-order perturbation theory is tunable by the frequency and amplitude of the driving field. 
We find the optimal frequency range to enhance the pairing interaction and show that the modified effective system can strengthen the two-body bound state. 
Our study suggests that external driving light can reinforce superconducting pair states in strongly correlated electron systems. 
\end{abstract}

\maketitle


\section{Introduction}
Manipulation of quantum materials utilizing light is one of the attractive research topics in condensed matter physics~\cite{Basov2017,Ishihara2019,delaTorre2021,Bloch2022}. 
Experimental endeavors such as light-induced phase transitions~\cite{Iwai2006,Giannetti2016,Koshihara2022} and nonlinear optical responses, including high-order harmonic generation~\cite{Ghimire2019,Shimano2020,Ma2021}, stimulate studies of optically driven quantum states of matter. 
To engineer periodically driven systems, Floquet theory provides useful methodologies~\cite{Bukov2015,Oka2019}. 
For example, Floquet formalism enables us to build an effective static Hamiltonian incorporating a light field, which can provide characteristics of a periodically driven quantum state. 
In a graphene system, an effective model coupled with circularly polarized light (CPL) predicts the emergence of a topological phase and a light-induced Hall effect~\cite{Oka2009,Kitagawa2011}.  
To date, the idea of Floquet engineering has been applied to various systems, including quantum magnets~\cite{Mentink2015,Sato2016,Kitamura2017,Claassen2017,Arakawa2021,Kobayashi2021,Shan2021} and superconductors~\cite{Benito2014,Takasan2017,Chono2020,Kitamura2016,Fujiuchi2020,Kennes2019,Kumar2021,Kitamura2022,Anan2024}. 

As for superconductors, optical control and induction of superconductivity are challenging research topics~\cite{Fausti2011,Mitrano2016,Buzzi2020,Sentef2016,Knap2016,Murakami2016,Wang2018,Claassen2019,Kaneko2019,Ejima2020,Murakami2022,Ueda2024,Gassner2024}. 
Floquet formalism has been applied to engineer electronic band structures of superconductors. 
Floquet engineering of topological features of superconductors has been studied in this context~\cite{Benito2014,Takasan2017,Chono2020}. 
Another application of Floquet formalism to superconductors is the control of effective interactions such as magnetic interactions that contribute to pairing glues in correlated electronic systems~\cite{Kitamura2016,Fujiuchi2020,Kennes2019,Kumar2021,Kitamura2022,Anan2024}. 
In repulsive Hubbard systems, light-induced $d$-wave superconductivity and possible transitions to topological superconductivity have been proposed via the derivation of effective $t$-$J$ models modified by periodic light fields~\cite{Kennes2019,Kumar2021,Kitamura2022,Anan2024}.  
In these systems, the parameters (e.g., frequency and amplitude) of the external fields become the control knobs of the effective models, which enable us to create renewed trends in superconductors. 

\begin{figure}[b]
\centering
\includegraphics[width=\columnwidth]{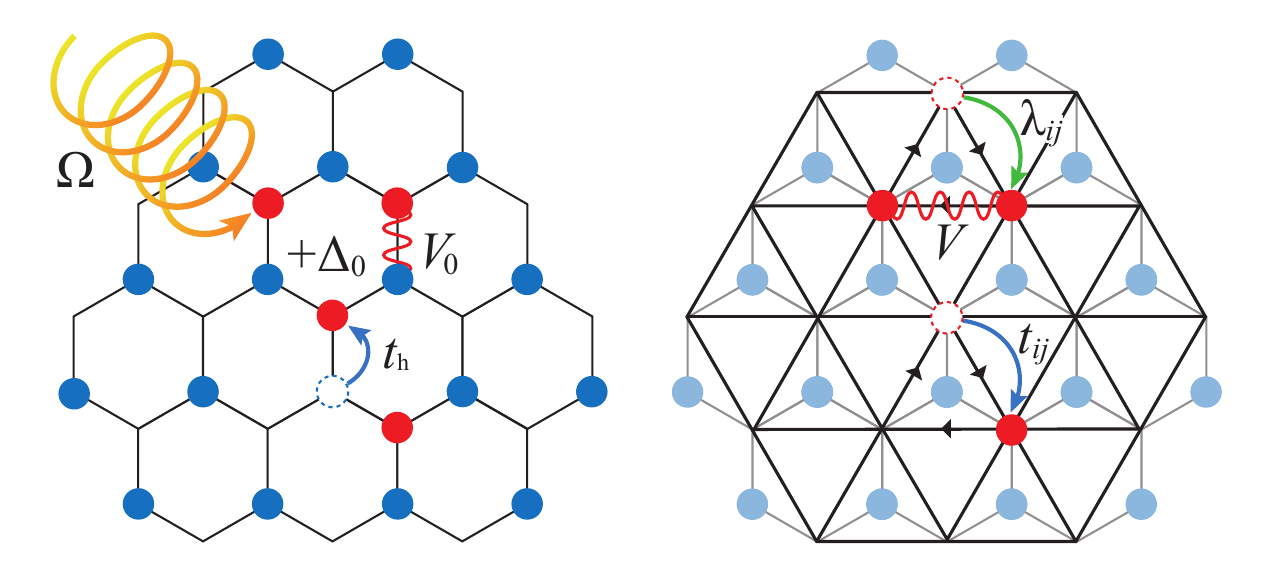}   
\caption{Left: interacting fermions on the honeycomb
lattice under CPL, where the energy level of the B sublattice is higher than that of the A sublattice. 
The blue and red circles represent fermions on the A and B sublattices, respectively. 
Right: effective
triangular system for doped fermions on the B sublattice. The light blue circles indicate the fermions on the A sublattice omitted in the effective model.}
\label{fig1}
\end{figure}

In this paper, we consider a carrier-doped system on a correlated band insulator~\cite{Crepel2021,Crepel2022prb,Crepel2022pnas,He2023,Crepel2023} as a target system for Floquet engineering of superconductivity. 
In the spinless fermion model on a honeycomb lattice with large sublattice potentials (see Fig.~\ref{fig1}), the system away from half filling can be effectively described by doped fermions on the higher-energy sublattice by projecting out the fully occupied sites in the lower-energy sublattice~\cite{Crepel2021}. 
As shown by Cr\'{e}pel and Fu~\cite{Crepel2021}, doped fermions interact with each other via perturbative excitation processes, where the intersublattice repulsion gives rise to the effective pairing interaction between doped fermions. 
Using the pairing glue obtained in the effective system, previous studies estimated the superconducting properties, such as pair binding energy, critical temperature, and pairing symmetry~\cite{Crepel2021,Crepel2022prb,Crepel2022pnas,He2023,Crepel2023}. 
One theoretical advantage of this system is that the attractive interaction is analytically obtainable with the strong-coupling expansion. 
This aspect is suitable for incorporating controllability afforded by external fields into effective pairing interactions using the perturbative expansion for periodically driven systems~\cite{Mentink2015,Kitamura2022}. 

From this perspective, we investigate the Floquet effective system of correlated spinless fermions on a honeycomb lattice under CPL (see Fig.~\ref{fig1}).  
We derive the effective model away from half filling, in which external CPL can tune the effective hoppings and interactions of fermions doped on the conduction network.  
Then, we present the model parameters as a function of the frequency and amplitude of the driving field. 
We find regions where CPL can encourage attractive interaction between doped fermions.  
Moreover, to demonstrate the signature of strong binding of doped fermions, we solve the two-body problem and show the strengthening of the bound state.    
Our finding suggests that light can generate an attractive contribution for pair formation in correlated electron systems.

The rest of this paper is organized as follows. 
In Sec.~\ref{sec.model}, we introduce the correlated spinless fermion model and derive the effective model under CPL obtained by the strong-coupling expansion. 
In Sec.~\ref{sec.eff}, we present the model parameters and discuss the nature of the effective Hamiltonian under CPL. 
In Sec.~\ref{sec.bou}, we show the results of the two-body problem to discuss the pairing properties.   
We discuss the reason for the enhanced pairing interaction in Sec.~\ref{sec.dis}.
Finally, in Sec.~\ref{sec.sum}, we summarize the findings of our work.


\section{Model}\label{sec.model}

We consider the correlated spinless fermion model on a honeycomb lattice~\cite{Crepel2021} described by
\begin{equation}
\hat{H} =  \hat{H}_{0} + \hat{T}_{\rm h},
\end{equation}
with
\begin{equation}
\hat{H_{0}} = V_{0}\sum_{\bm{r}\in \rm A}\sum_{j=1,2,3}\hat{n}_{\bm{r}}^{\rm A}\hat{n}_{\bm{r}+\bm{d}_{j}}^{\rm B} 
+ \Delta_{0}\sum_{\bm{r}\in \rm B}\hat{n}_{\bm{r}}^{\rm B}
\end{equation}
and 
\begin{equation}
\hat{T}_{\rm h} =-t_{\rm h}\sum_{\bm{r}\in \rm A}\sum_{j=1,2,3}\left( \hat{b}_{\bm{r}+\bm{d}_{j}}^{\dagger}\hat{a}_{\bm{r}} + {\rm H.c.}  \right),
\label{eq(3)} 
\end{equation}
where $\hat{a}_{\bm{r}}$ ($\hat{a}^{\dag}_{\bm{r}}$) and $\hat{b}_{\bm{r}}$ ($\hat{b}^{\dag}_{\bm{r}}$) are the annihilation (creation) operators for fermions on the A and B sites, respectively, and $\hat{n}_{\bm{r}}^{\rm A}=\hat{a}^{\dag}_{\bm{r}}\hat{a}_{\bm{r}}$ and $\hat{n}_{\bm{r}}^{\rm B}=\hat{b}^{\dag}_{\bm{r}}\hat{b}_{\bm{r}}$.
$\bm{r}$ represents the position on the honeycomb lattice, and $\bm{d}_j$ gives the position of the B site relative to the A site, where $\bm{d}_{1}=d(0,1)$, $\bm{d}_{2}= d(\sqrt{3}/2,-1/2)$, and $\bm{d}_{3}= d(-\sqrt{3}/2,-1/2)$. 
$V_0$~$(>0)$ is the repulsive interaction between the nearest-neighboring fermions, and $\Delta_0$~$(>0)$ is the energy-level differences between the A and B sublattices.
$t_{\rm h}$ is the hopping integral between the adjacent A and B sites. 

We address the model doped away from half filling. 
When $t_{\rm h}=0$, the A sublattice is fully occupied ($ \braket{\hat{n}^{\rm A}_{\bm{r}}}=1$), and the doped fermions partially occupy the B sublattice. 
In the band picture, the ground state at half filling is a band insulator, and fermions doped on it reside in the conduction band. 
When $\Delta_0 \gg t_{\rm h}$, we can approximately project out the A-sublattice degrees of freedom, and the model reduces into an effective itinerant-fermion model on the triangular network of the B sublattice (see Fig.~\ref{fig1})~\cite{Crepel2021}.   
The effective model becomes an interacting fermion model when $V_0>0$. 
As shown in Ref.~\cite{Crepel2021}, the repulsive $V_0$ can lead to attractive interactions between fermions and create a superconducting state in the effective B sublattice system. 

To manipulate the effective pairing interactions using light, we derive an effective model under a periodic driving field.  
Here, we consider time-periodic CPL, whose vector potential is given by 
\begin{align} 
\bm{A}(t) = \frac{E_{0}}{\Omega}\left( \sin{\Omega t}, \cos{\Omega t} \right), 
\label{eq:vec_CPL}
\end{align} 
where $E_0$ is the amplitude of the electric field and $\Omega$ is the frequency. 
The electric field is given by $\bm{E}(t)=-\partial_t \bm{A}(t)$. 
This external field is introduced via the Peierls substitution: \begin{equation}
\hat{T}_{\rm h}(t) =-t_{\rm h}\sum_{\bm{r} \in \rm A}\sum_{j=1,2,3}\left( e^{-i\frac{e}{\hbar}\bm{A}(t)\cdot \bm{d}_j}\hat{b}_{\bm{r}+\bm{d}_{j}}^{\dagger}\hat{a}_{\bm{r}} + {\rm H.c.}  \right),
\label{eq:ham_A}
\end{equation}
where $e$ $(>0)$
is the electron charge and $\hbar$ is the reduced Planck constant. 
Using the Jacobi-Anger expansion, we can transform this time-periodic Hamiltonian into
\begin{align} 
\hat{T}_{\rm h}(t) = \sum_{m=-\infty}^{\infty}\hat{T}_{m}e^{-im\Omega t},
\end{align} 
with
\begin{align} 
\hat{T}_{m} = -t_{\rm h}\sum_{\bm{r} \in \rm A} \sum_{j=1,2,3}
&i^{m}J_{m}(\mathcal{A})e^{-im\alpha_{j}}
\notag \\
\times & \left[(-1)^{m}\hat{b}_{\bm{r}+\bm{d}_{j}}^{\dagger}\hat{a}_{\bm{r}} + \hat{a}_{\bm{r}}^{\dag}\hat{b}_{\bm{r}+\bm{d}_{j}}\right],\label{eq(6)}
\end{align} 
where $\alpha_{1}=0$, $\alpha_{2}=-2\pi/3$, and $\alpha_{3}=2\pi/3$. 
$J_{m}(\mathcal{A})$ is the $m$th-order Bessel function of the first kind, and $\mathcal{A}={eE_{0}d/\hbar\Omega}$ corresponds to the dimensionless Floquet parameter~\cite{delaTorre2021,Mentink2015}. 

We derive an effective model using the transformation $\hat{H}_{\rm eff} = e^{i\hat{\Lambda}(t)} [ \hat{H}(t) - i \partial_t ]e^{-i\hat{\Lambda}(t)}$~\cite{Kitamura2022}. 
We present the details of the derivation of our effective model in Appendix~\ref{app:der}. 
We conduct the strong-coupling expansion up to the second order
assuming that $t_{\rm h}$ is much smaller than the energy scale of $\hat{H}_0$, and the frequency $\Omega$ is off-resonant.  
Under this condition, we can make an effective model projecting out the fermions fully occupying the A sublattice. 
The second-order perturbative processes caused by the hopping $t_{\rm h}$ lead to the effective interactions among the doped fermions on the B sublattice. 
There are three types of excited states in the perturbative processes, which are classified by $n^{\rm B}_{\triangle}$, the number of occupied B sites on an upper triangle (three sites) around an A site~\cite{Crepel2021,DPT}.
The energies of these excited states are given by $\Delta_0$, $\Delta_0+V_0$, and $\Delta_0+2V_0$ for $n^{\rm B}_{\triangle}=$ 2, 1, and 0, respectively.  

We incorporate the effects of CPL into these perturbative processes and derive an effective model (see Appendix~\ref{app:der} for details). 
The resulting effective Hamiltonian on the triangular lattice of the B sites is given by 
\begin{align}
\hat{H}_{\rm eff} & = 
\sum_{\langle i,j \rangle} \left( t_{ij} \hat{b}^{\dag}_i \hat{b}_j + {\rm H.c.} \right)
+\sum_{\langle i,j,k\rangle \in \triangle} 
\left( \lambda_{ij} \hat{b}^{\dag}_i \hat{n}_k \hat{b}_j + {\rm H.c.} \right)
\notag \\
&+ V \sum_{\langle i,j\rangle} \hat{n}_{i}\hat{n}_{j}
+U_3 \sum_{\langle i,j,k\rangle \in \triangle} \hat{n}_{i}\hat{n}_{j}\hat{n}_{k}, 
\label{eq:Heff}
\end{align}
where $\hat{b}_{j}$ ($\hat{b}^{\dag}_{j}$) is the fermion annihilation (creation) operator at site $j$ on the B sublattice and $\hat{n}_{j}=\hat{b}^{\dag}_{j}\hat{b}_{j}$. 
$\langle i,j \rangle$ indicates a pair of nearest-neighbor sites on the B sublattice, and $\langle i,j,k\rangle \in \triangle$ indicates three nearest-neighbor B sites around a filled A site. 
$t_{ij}$ is the effective hopping of a fermion between nearest-neighbor B sites, given by 
\begin{align}
t_{ij} & = \sum_{m=-\infty}^{\infty}
\frac{t_{\rm{h}}^2 J_{m}(\mathcal{A})^2 }{\Delta_{0}+V_{0}-m \hbar \Omega}e^{i \nu_{ij} \frac{2\pi m}{3}}, 
\label{eq:tf}
\end{align}
where $\nu_{ij} = \pm 1$ is the sign characterizing the rotation (clockwise or counterclockwise) direction of the hopping (see Fig.~\ref{fig1}). 
$\lambda_{ij}$ is the correlated hopping, given by 
\begin{align}
\lambda_{ij} &=\sum_{m=-\infty}^{\infty} \left(
\frac{t_{\rm{h}}^2 J_{m}(\mathcal{A})^2 }{\Delta_{0}-m \hbar \Omega}
-\frac{t_{\rm{h}}^2 J_{m}(\mathcal{A})^2 }{\Delta_{0}+V_{0}-m \hbar \Omega} 
\right)
e^{i \nu_{ij} \frac{2\pi m}{3}}. 
\label{eq:lambda}
\end{align}
$\lambda_{ij}$ is active when two of three sites on a triangle are occupied.  
$V$ is the two-body interaction, given by 
\begin{align}
V &= \sum_{m=-\infty}^{\infty}
\Biggr(
-\frac{ t_{\rm{h}}^2 J_{m}(\mathcal{A})^2 }{\Delta_{0}-m \hbar \Omega}
 + \frac{ 4t_{\rm{h}}^2 J_{m}(\mathcal{A})^2 }{\Delta_{0}+V_{0}-m \hbar \Omega}
\notag\\            
&\hspace{45pt}
-\frac{ 3t_{\rm{h}}^2 J_{m}(\mathcal{A})^2 }{\Delta_{0}+2V_{0}-m \hbar \Omega}
\Biggl),
\label{eq:V}
\end{align}
and $U_3$ is the three-body interaction, given by 
\begin{align}
U_{3} &=  \sum_{m=-\infty}^{\infty}
\Biggr(
\frac{ 3t_{\rm{h}}^2 J_{m}(\mathcal{A})^2 }{\Delta_{0}-m \hbar \Omega}
- \frac{ 6t_{\rm{h}}^2 J_{m}(\mathcal{A})^2 }{\Delta_{0}+V_{0}-m \hbar \Omega} 
\notag \\
&\hspace{45pt}
+ \frac{ 3t_{\rm{h}}^2 J_{m}(\mathcal{A})^2 }{\Delta_{0}+2V_{0}-m \hbar \Omega}
\Biggl). 
\label{eq:U3}
\end{align}
Due to the effect of CPL, the denominators of the effective parameters include $m\hbar\Omega$.  
Note that we omit the energy term, which gives the constant mainly determined by $\hat{H}_0$, from the effective Hamiltonian $\hat{H}_{\rm eff}$ in Eq.~(\ref{eq:Heff}) (see Appendix~\ref{app:der}). 

Here, we comment on the validity of the effective Hamiltonian.  
When $m\hbar\Omega = \Delta_{0} + qV_{0}$ ($q=0,1,2$), i.e., an energy gap of $\hat{H}_0$ is equal to an energy of a multiphoton process, the effective parameters in Eqs.~(\ref{eq:tf})--(\ref{eq:U3}) diverge, indicating that the perturbatively obtained effective Hamiltonian is not valid.  
In this case, excitation is resonant, where the system absorbs light substantially. 
We should also note the thermalization problem~\cite{delaTorre2021}. 
Because we consider an isolated system, the system during CPL irradiation keeps absorbing energy and finally heats up to a featureless infinite-temperature state. 
This heating is fast when $\Omega$ is a resonant frequency, at which the effective interactions diverge.
In other words, slow heating can be expected when $\Omega$ is set to an off-resonant frequency~\cite{delaTorre2021}. 
In general, there is a prethermal plateau before the infinite-temperature ensemble is reached, and effective Floquet Hamiltonians are valid on the prethermal time scale~\cite{delaTorre2021,Mori2016,Mori2023,Bukov2016,Abanin2017,Kuwahara2016}. 
Thus, our effective Floquet state is valid in the prethermal regime when $\Omega$ is set to an off-resonant frequency.


\section{Effective hoppings and interactions}\label{sec.eff}

\begin{figure}[b]
\begin{center}
\includegraphics[width=0.95\columnwidth]{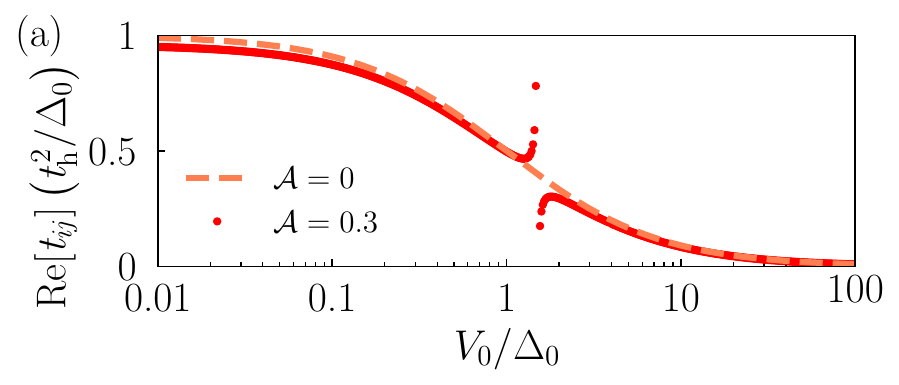}   
\\
\includegraphics[width=0.95\columnwidth]{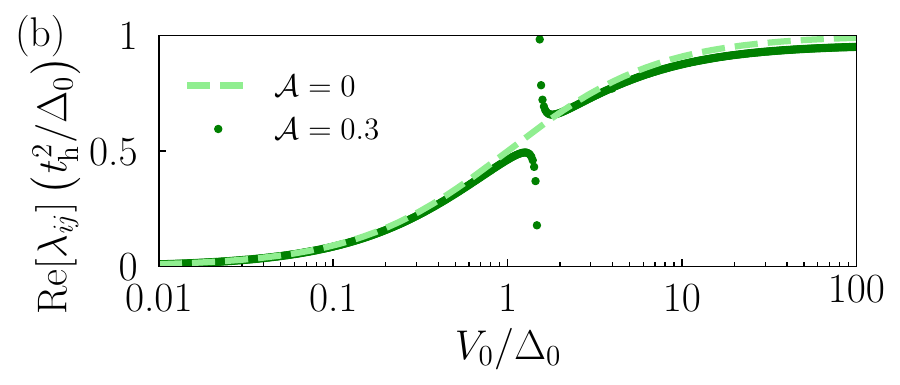}
\\
\includegraphics[width=0.95\columnwidth]{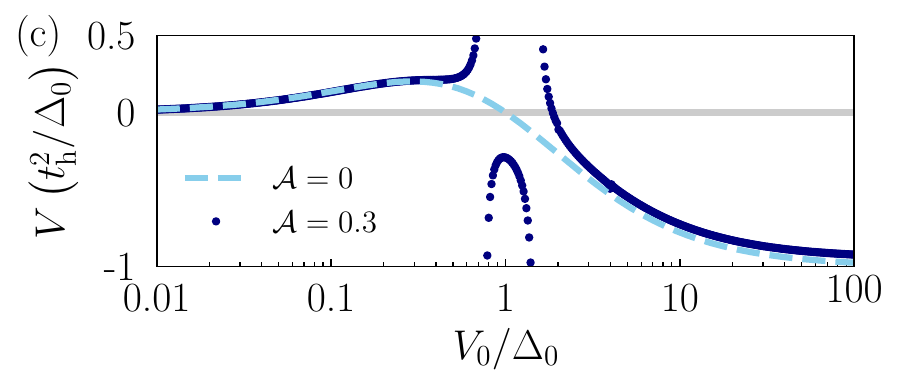}   
\\
\includegraphics[width=0.95\columnwidth]{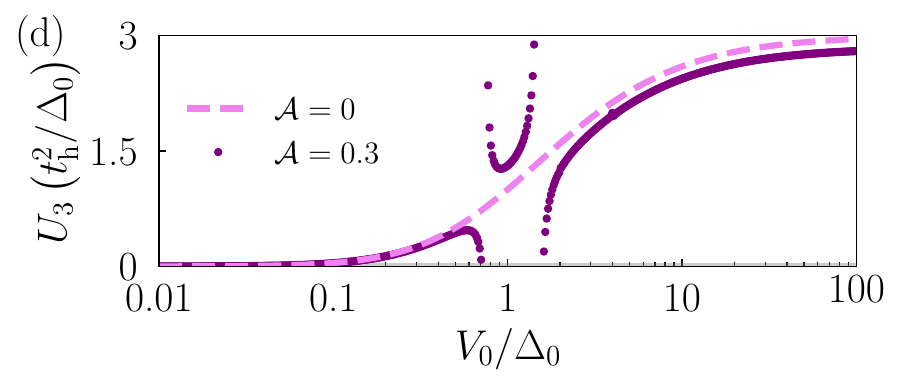}
\caption{$V_0$ dependences of (a) $t_{ij}$, (b) $\lambda_{ij}$, (c) $V$, and (d) $U_{3}$ for $\mathcal{A}=0$ and  $\mathcal{A}=0.3$ with $\hbar\Omega/\Delta_0 = 2.5$. 
The real parts of $t_{ij}$ and $\lambda_{ij}$ are plotted in (a) and (b), respectively.} 
\label{fig2}
\end{center}
\end{figure}

First, we review the behaviors of the model parameters without CPL.  
In Fig.~\ref{fig2}, we show the effective hoppings and interactions as a function of $V_0 / \Delta_0$. 
Their $V_0$ dependences are consistent with those in Ref.~\cite{Crepel2021}.
When $V_0=0$, $t_{ij} = t_{\rm{h}}^2 / \Delta_0$ (at $\mathcal{A}=0$), and $\lambda_{ij}=V=U_3=0$; i.e., the system is described by free fermions. 
As seen in Figs.~\ref{fig2}(a) and \ref{fig2}(b), the amplitudes of $t_{ij}$ are suppressed, whereas $\lambda_{ij}$ increases as $V_0$ increases. 
Their amplitudes are reversed at $V_{0}/\Delta_{0}=1$. 
The free hopping $t_{ij}$ is dominant in the weak-coupling region, where $t_{ij} \simeq t_{\rm{h}}^2 / \Delta_0$ and $\lambda_{ij} \simeq (t_{\rm{h}} / \Delta_0)^2 V_0 \rightarrow 0$ at $V_{0} \rightarrow 0$. 
On the other hand, the correlated hopping $\lambda_{ij}$ is dominant in the strong-coupling region, where $t_{ij} \simeq t_{\rm{h}}^2/V_0 \rightarrow 0$ and $\lambda_{ij} \simeq t_{\rm{h}}^2 / \Delta_0$ at $V_{0} \rightarrow \infty$.  
Meanwhile, the effective two-body interaction $V$ is activated by $V_0$ and turns from repulsion to attraction at $V_0/\Delta_0=1$ [see Fig.~\ref{fig2}(c)]. 
$V \simeq 2 (t_{\rm h} / \Delta_0)^2 V_0> 0$ at $V_0/\Delta_0 \ll 1$, whereas $V \simeq -t_{\rm h}^2 / \Delta_0 < 0$ at $V_0/\Delta_0 \gg 1$.  
Hence, we can obtain an effective attracting interaction $V<0$ mediated by the repulsive interaction $V_0>0$.  
As shown in Fig.~\ref{fig2}(d), $U_3$ monotonically increases as $V_0$ increases. 
At $V_0/\Delta_0 \ll 1$, $U_3 \propto t_{\rm h}^2V_0^2/\Delta^3_0$ ($\ll V, \lambda_{ij}$) is negligible, and thus, the effective system is described by nearly free fermions weakly correlated via $V$ and $\lambda_{ij}$. 
At $V_0/\Delta_0 \gg 1$, $U_3 \simeq 3t_{\rm h}^2/\Delta_0$ does not favor fully occupied triangle units, whereas pairs of fermions can be formed via $V$ and $\lambda_{ij}$. 
In the model without CPL, a previous study~\cite{Crepel2021} showed that the pairing glue is characterized by $V-2\lambda_{ij}$ ($<0$), and the binding energy of a pair becomes nonzero at $V_0>0$~\cite{Crepel2021}. 
The pairing interaction increases as $V_0/\Delta_0$ increases; i.e., the pairing state smoothly crosses over between the BCS and Bose-Einstein condensation regimes as a function of $V_0$~\cite{Crepel2021}. 

\begin{figure}[t]
\centering
\includegraphics[width=0.95\columnwidth]{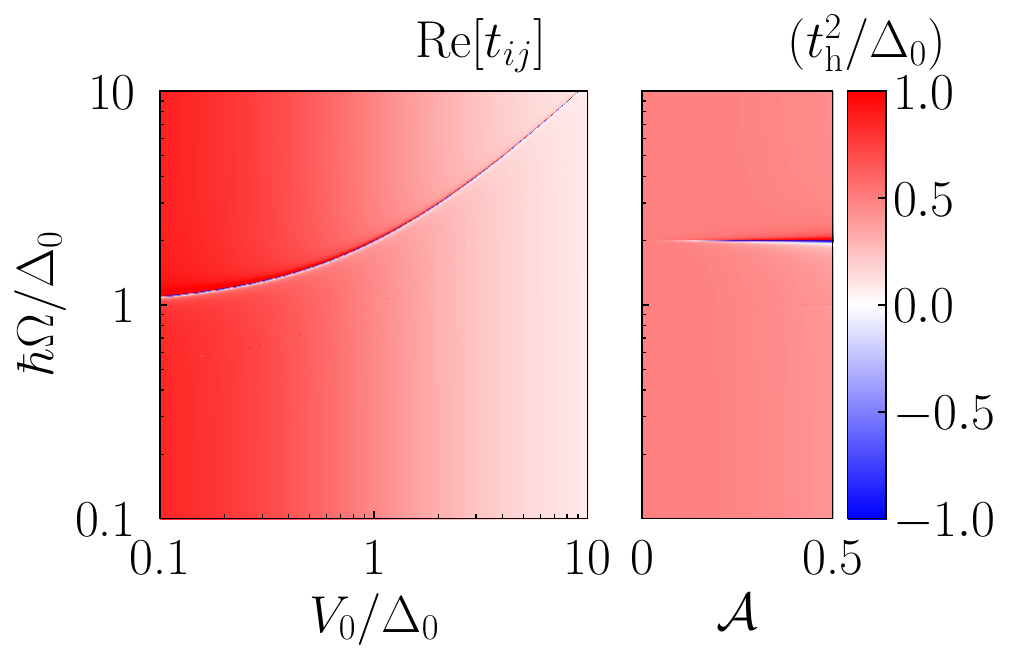}   
\\
\includegraphics[width=0.95\columnwidth]{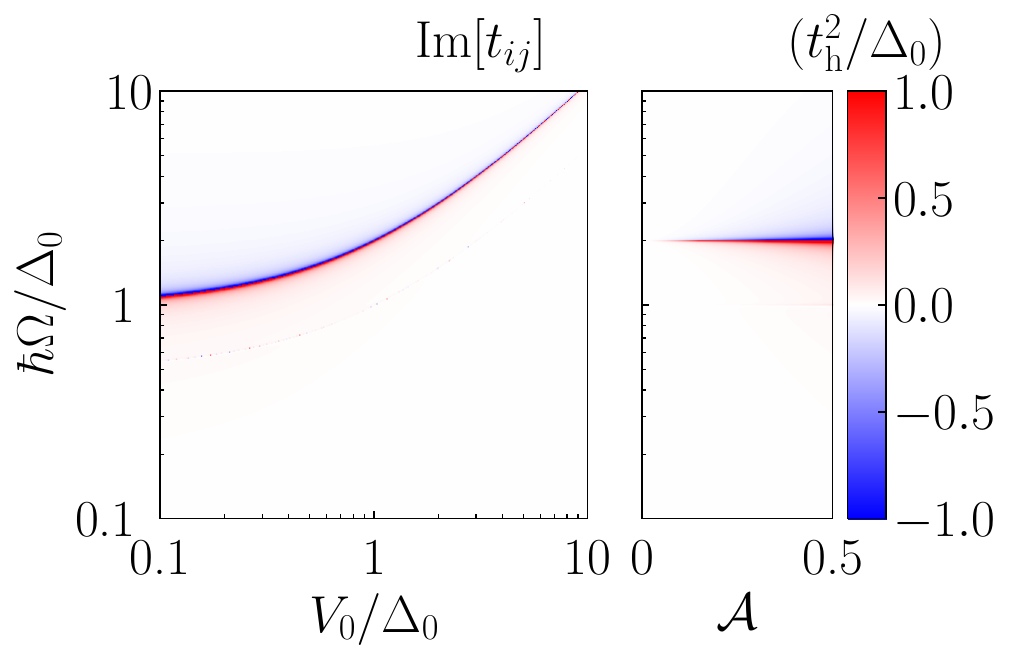}
\caption{Left: real and imaginary parts of $t_{ij}$ in the $V_0$-$\Omega$ plane at $\mathcal{A}=0.3$. 
Right: real and imaginary parts of $t_{ij}$ in the $\mathcal{A}$-$\Omega$ plane at $V_0/\Delta_0=1$.} 
\label{fig3}
\end{figure}

\begin{figure}[t]
\centering
\includegraphics[width=0.95\columnwidth]{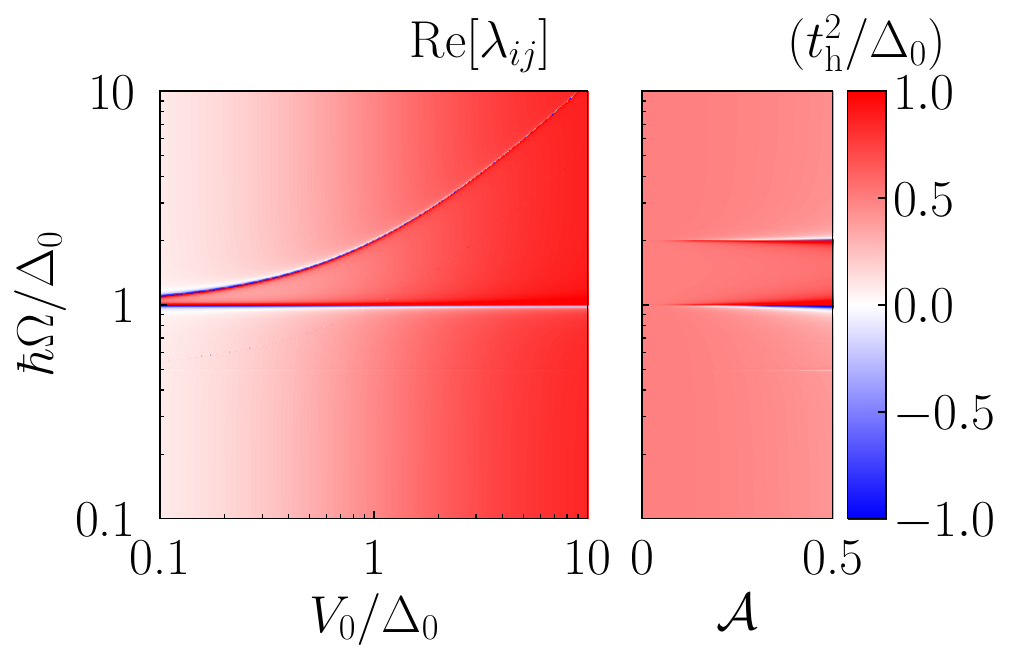}   
\\
\includegraphics[width=0.95\columnwidth]{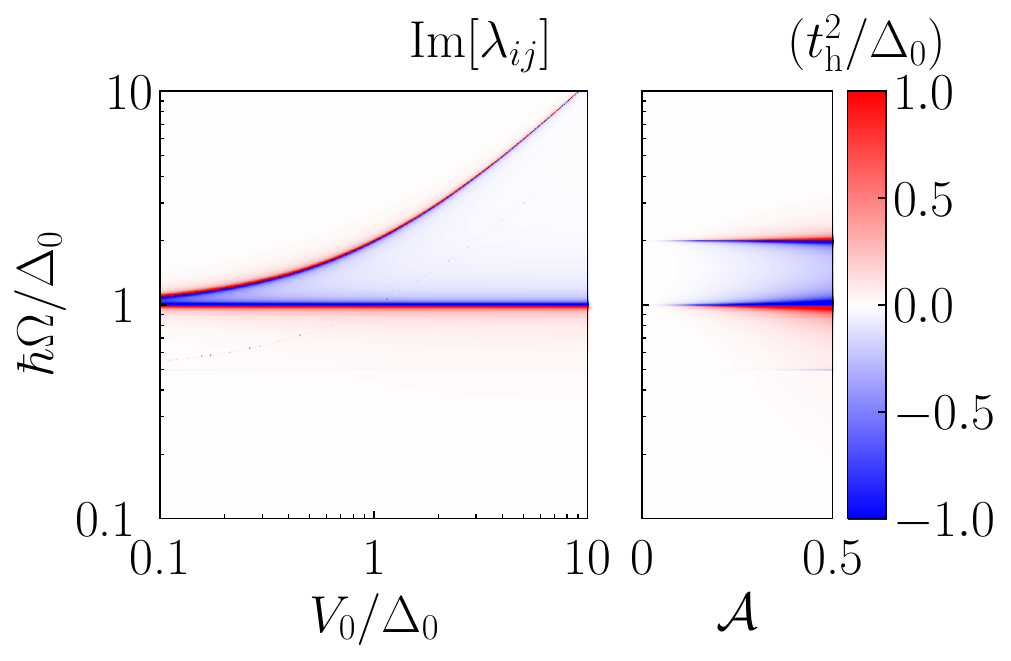}
\caption{Left: real and imaginary parts of $\lambda_{ij}$ in the $V_0$-$\Omega$ plane at $\mathcal{A}=0.3$. 
Right: real and imaginary parts of $\lambda_{ij}$ in the $\mathcal{A}$-$\Omega$ plane at $V_0/\Delta_0=1$.} 
\label{fig4}
\end{figure}

\begin{figure}[t]
\centering
\includegraphics[width=0.95\columnwidth]{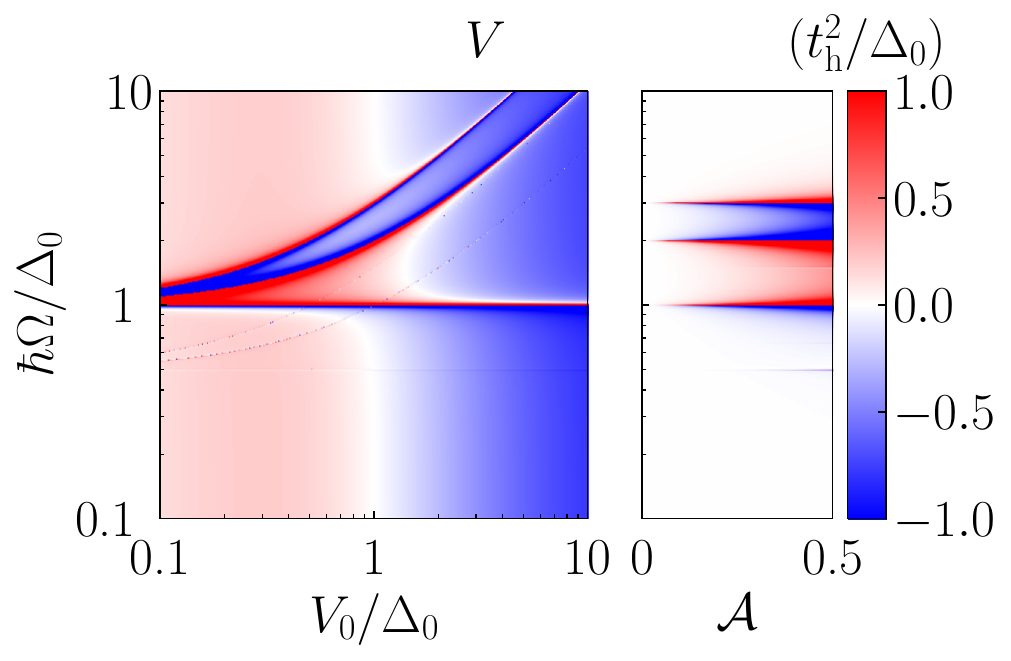}
\caption{Left: $V$ in the $V_0$-$\Omega$ plane at $\mathcal{A}=0.3$. 
Right: $V$ in the $\mathcal{A}$-$\Omega$ plane at $V_0/\Delta_0=1$.} 
\label{fig5}
\end{figure}

\begin{figure}[t]
\centering
\includegraphics[width=0.9\columnwidth]{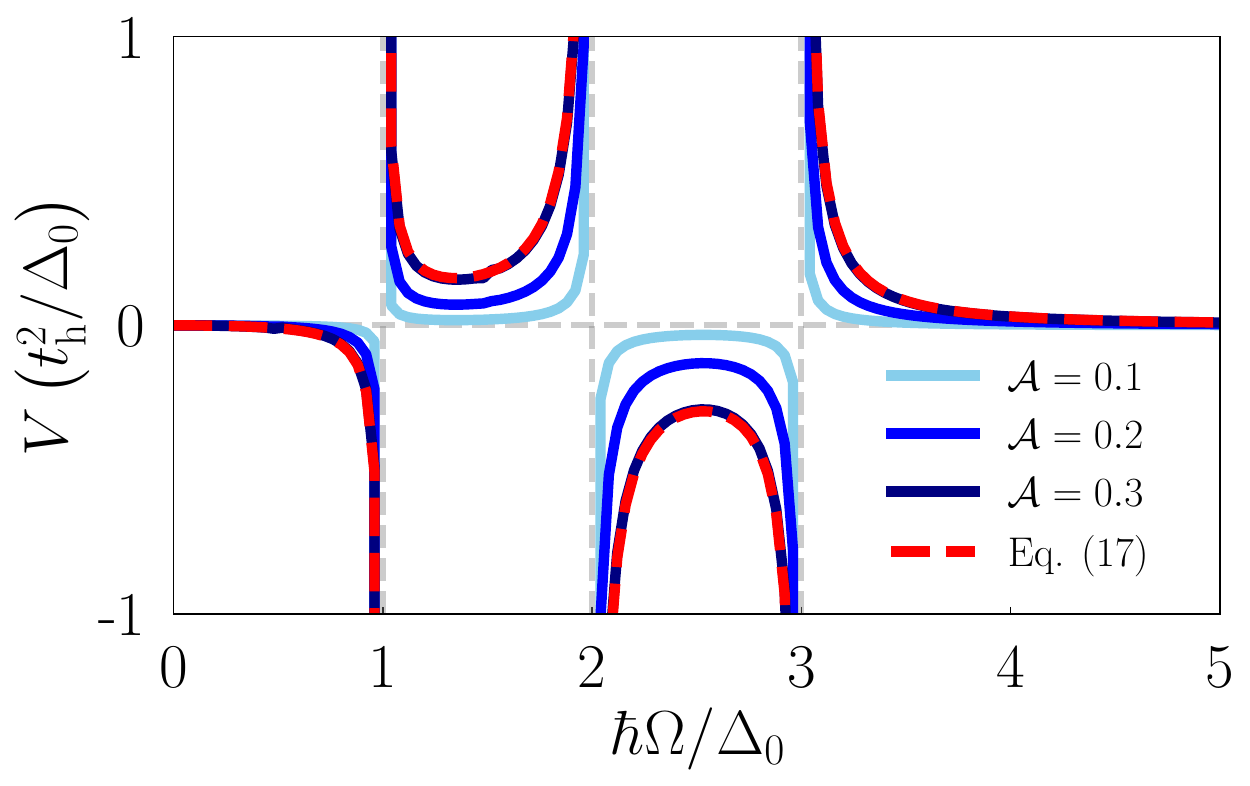}   
\caption{$V$ as a function of $\hbar\Omega/\Delta_0$ at $V_0/\Delta_0=1$, where $V_{\mathcal{A}=0}=0$. 
$V$ in Eq.~(\ref{eq:V_weakA}) is compared to $V$ in Eq.~(\ref{eq:V}) at $\mathcal{A}=0.3$.} 
\label{fig6}
\end{figure}

\begin{figure}[t]
\centering
\includegraphics[width=0.95\columnwidth]{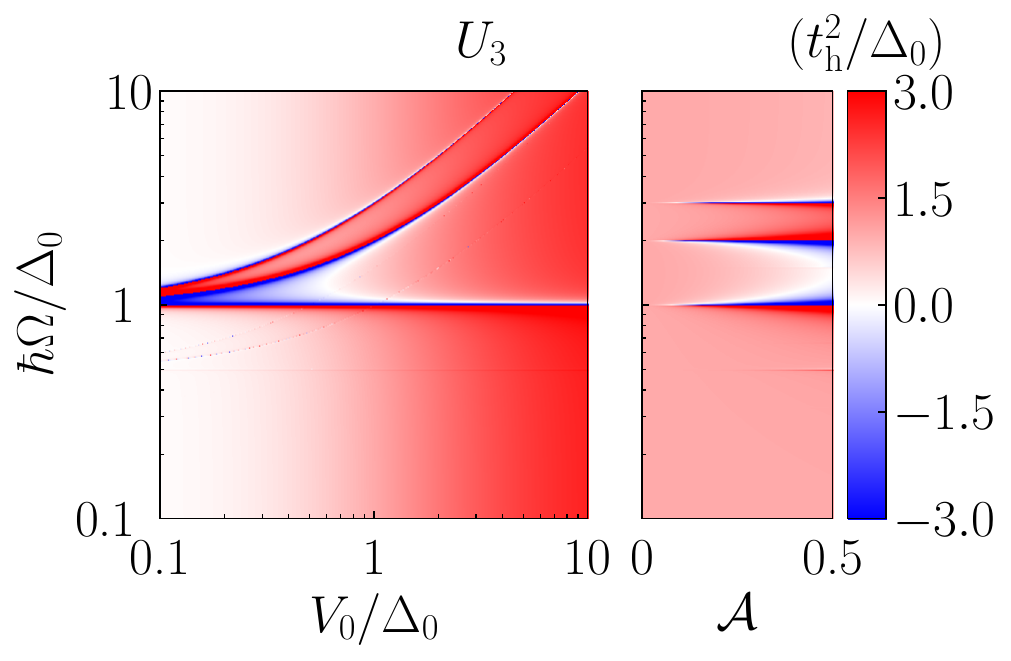}
\caption{Left: $U_3$ in the $V_0$-$\Omega$ plane at $\mathcal{A}=0.3$. 
Right: $U_3$ in the $\mathcal{A}$-$\Omega$ plane at $V_0/\Delta_0=1$.} 
\label{fig7}
\end{figure}

Let us consider the effects of CPL. 
The effective hoppings and interactions depending on $\mathcal{A}$ and $\Omega$ are presented in Figs.~\ref{fig3}--\ref{fig7}. 
In Fig.~\ref{fig2}, we also present the effective parameters at $\mathcal{A}=0.3$ and $\hbar\Omega=2.5$. 
We show the real and imaginary parts of $t_{ij}$ in Fig.~\ref{fig3}. 
$t_{ij}$ decreases as $V_0$ increases, reflecting the property of the model without CPL, as seen in Fig.~\ref{fig2}(a).   
Corresponding to Eq.~(\ref{eq:tf}), $t_{ij}$ shows a divergence at $\hbar\Omega = \Delta_0 + V_0$ [see Fig.~\ref{fig2}(a)]. 
Our effective model derived from the second-order perturbation theory is invalid at this divergent point.   
Although $t_{ij}$ at $\mathcal{A}=0$ is real, $t_{ij}$ at $\mathcal{A}\ne 0$ possesses an imaginary part. 
As seen in Eq.~(\ref{eq:tf}), the imaginary part ${\rm Im} \left[ t_{ij} \right]$ is due to the phase factor $e^{\pm i(2\pi m/3)}$ attributed to CPL ($|m| \ge 1$). 
However, as shown in Fig.~\ref{fig3}, the imaginary part is much smaller than the real part (except for the divergent points). 
The lowest-order contribution to ${\rm Im} \left[ t_{ij} \right]$ is given by 
\begin{align}
{\rm Im} \left[ t_{ij} \right] 
\simeq \frac{t_{\rm h}^2 \hbar\Omega}{ 2\left[ (\Delta_{0} + V_{0} )^{2} - (\hbar\Omega)^{2} \right]} \mathcal{A}^2 \sin \frac{2\pi}{3} \nu_{ij}. 
\label{eq:tf_small_A}
\end{align}
This equation indicates that the imaginary part is generated by $\mathcal{A}$. 
This ${\rm Im} \left[ t_{ij} \right]$ vanishes at $\Omega \rightarrow 0$. 
On the other hand, ${\rm Im} \left[ t_{ij} \right] \propto (\hbar\Omega)^{-1}$ at $\hbar\Omega \gg \Delta_0+V_0$ decreases as $\Omega$ increases. 
Corresponding to Eq.~(\ref{eq:tf_small_A}), ${\rm Im} \left[ t_{ij} \right]$ changes sign at $\hbar\Omega = \Delta_0+V_0$ (see Fig.~\ref{fig3}). 
The complex hopping $t_{ij}$ resulting from CPL is similar to the next-nearest-neighbor hopping in the Haldane model on a honeycomb lattice~\cite{Haldane1988}, which yields alternating flux with broken time-reversal symmetry. 
Different from the high-frequency expansion of the graphene system~\cite{Oka2009}, the sublattice potential $\Delta_0$ and the interaction $V_0$ are taken into account in Eq.~(\ref{eq:tf_small_A}). 

In Fig.~\ref{fig4}, we present the real and imaginary parts of the correlated hopping $\lambda_{ij}$. 
Reflecting the zero-field property shown in Fig.~\ref{fig2}(b), $\lambda_{ij}$ increases as $V_0$ increases. 
Corresponding to $\lambda_{ij}$ in Eq.~(\ref{eq:lambda}), $\lambda_{ij}$ diverges at $\hbar \Omega = \Delta_0$ and $\hbar \Omega = \Delta_0 + V_0$.  
$\lambda_{ij}$ also possesses an imaginary part, although its value is small compared with the real part. 
${\rm Im}[\lambda_{ij}]$ shows sign changes at $\hbar \Omega = \Delta_0$ and $\hbar \Omega = \Delta_0 + V_0$. 
This behavior can be understood from the lowest-order contribution to ${\rm Im} \left[ \lambda_{ij} \right]$, given by 
\begin{align} 
{\rm Im} \left[ \lambda_{ij} \right] 
\simeq 
\frac{t_{\rm h}^2 \hbar\Omega V_{0}(2\Delta_0 \! + \! V_{0} )}{2\left[ \Delta_{0}^{2} \!-\! (\hbar \Omega)^{2} \right]\left[ (\Delta_{0} \!+\! V_{0} )^{2} \!-\! (\hbar\Omega)^{2}\right]}
\mathcal{A}^2  \sin \frac{2\pi}{3} \nu_{ij}. 
\label{eq:lambda_small_A}
\end{align}
The $\Omega$ dependence of this equation is consistent with the sign change of ${\rm Im} \left[ \lambda_{ij} \right]$ at $\hbar\Omega = \Delta_0$ and $\hbar\Omega = \Delta_0+V_0$ seen in Fig.~\ref{fig4}. 
      
Figure~\ref{fig5} shows the two-body interaction $V$ for doped fermions. 
Corresponding to $V$ without CPL, $V$ changes sign around $V_0 / \Delta_0=1$; i.e., $V$ becomes attractive in the large-$V_0$ region. 
Interestingly, $V$ under CPL becomes negative at $\Delta_0 + V_0 < \hbar\Omega < \Delta_0 + 2V_0$ even though $V$ without the external field is zero or positive at $V_0/\Delta_0 \le 1$. 
This suggests that $V$ can be attractive by applying CPL.  

This attractive interaction is caused by the terms including $\mathcal{A}$ in Eq.~(\ref{eq:V}). 
When $\mathcal{A} \ll 1$, the change in the effective interaction ($\delta V=V-V_{\mathcal{A}=0}$) is given by  
\begin{align}
\delta V &\simeq 
-\frac{1}{2} V_{\mathcal{A}=0} \mathcal{A}^2 
-\frac{1}{2}\Bigg[ \frac{t_{\rm h}^2\Delta_{0}}{ \Delta_{0}^2 \!-\! (\hbar \Omega)^2} 
\notag \\
&-\frac{4t_{\rm h}^2(\Delta_{0}+V_0)}{(\Delta_{0}+V_0)^2 \!-\! (\hbar \Omega)^2} 
+\frac{3t_{\rm h}^2(\Delta_{0}+2V_0)}{(\Delta_{0}+2V_0)^2 \!-\! (\hbar \Omega)^2} 
\Bigg] 
\mathcal{A}^2 , 
\label{eq:dV_2nd}
\end{align}
where 
\begin{align}
V_{\mathcal{A}=0} 
&= -\frac{ t_{\rm{h}}^2  }{\Delta_{0}}
 + \frac{ 4t_{\rm{h}}^2}{\Delta_{0}+V_{0}}
-\frac{ 3t_{\rm{h}}^2 }{\Delta_{0}+2V_{0}}
\end{align}
is the effective interaction at $\mathcal{A}=0$. 
The first term in Eq.~(\ref{eq:dV_2nd}), $-V_{\mathcal{A}=0} \mathcal{A}^2/2$, indicates that the repulsive (attractive) interaction is suppressed by $\mathcal{A}$ when $V_{\mathcal{A}=0}$ is repulsive (attractive). 
This is a consequence of the reduction of the amplitude of the $m=0$ term due to the zeroth-order Bessel function, i.e., $|V_{\mathcal{A}=0}| J_{m=0}(\mathcal{A})^2 \le |V_{\mathcal{A}=0}|$. 
The remaining quadratic terms in Eq.~(\ref{eq:dV_2nd}) stem from the $m=\pm 1$ contributions, and they change sign depending on the frequency $\Omega$.  

To clarify the $\Omega$-dependent behavior, we consider the effective interaction $V$ at $V_0 = \Delta_0$, where $V_{\mathcal{A}=0}=0$.  
When $V_0=\Delta_0$, the lowest-order contribution to $V = \delta V$ is given by 
\begin{align}
V \simeq 
-\frac{t_{\rm h}^2\Delta_{0}(\hbar\Omega)^{2} \left[11\Delta_{0}^{2} \! +\! (\hbar\Omega)^{2}\right] }{\left[ \Delta_{0}^2 \!-\! (\hbar \Omega)^2\right]\left[(2\Delta_{0})^2 \!-\! (\hbar \Omega)^2\right]\left[(3\Delta_{0})^2 \!-\! (\hbar \Omega)^2\right]} \mathcal{A}^2. 
\label{eq:V_weakA}
\end{align}
Since the numerator is positive, the sign of $V$ depends only on $\Omega$ in the denominator. 
This formula indicates that $V$ is repulsive at $\Delta_0<\hbar\Omega<2\Delta_{0}$ and $3\Delta_0 <\hbar\Omega$, whereas $V$ is attractive at $0<\hbar\Omega<\Delta_{0}$ and $2\Delta_0 < \hbar\Omega < 3\Delta_0$. 
In Fig.~\ref{fig6}, we plot $V$ at $V_0=\Delta_0$ as a function of $\Omega$, where $V$ in Eq.~(\ref{eq:V_weakA}) shows good agreement with $V$, including all contributions in Eq.~(\ref{eq:V}) even at $\mathcal{A}=0.3$. 
As seen in Fig.~\ref{fig6}, the attractive (repulsive) $V$ emerges continuously from zero as $\mathcal{A}$ increases.  
The strength of the attraction $|V|$ is small at $0<\hbar\Omega<\Delta_{0}$, and $2\Delta_{0}<\hbar\Omega<3\Delta_{0}$ may be a promising region for the attractive interaction generated by CPL.  
Similar to the analytical estimation at $V_0=\Delta_0$, the effective interaction $V$ shown in Fig.~\ref{fig5} (left panel) obtains an attractive contribution at $\Delta_0 + V_0 < \hbar\Omega < \Delta_0 + 2V_0$. 
Note that the effective interaction obtained with the perturbation theory is valid when $V$ is on the order of $t^2_{\rm h}/\Delta_0$. For example, the strong-coupling expansion is valid when $V_0 \sim \Delta_0$ and $\hbar\Omega$ is set around the middle of the gap (i.e., $\hbar\Omega \simeq \Delta_0 + 1.5 V_0$).  
However, the $V_0 \ll \Delta_0$ region, where the gap between $\Delta_0+V_0$ and $\Delta_0+2V_0$ is small, is unfavorable for our effective theory.  

The three-body interaction $U_3$ is presented in Fig.~\ref{fig7}. 
When $\mathcal{A}$ is small, $U_3$ is repulsive in most regions. 
Hence, the formation of a three-body complex is hampered by $U_3$ even under CPL. 
$U_3$ suggests that, although the pair formation is promoted by an attractive $V$, pairs are uniformly distributed without configuring the phase separation of the carrier density.


\section{Bound state}\label{sec.bou}

We assess the pairing property by solving the two-body problem. 
A configuration of two fermions is given by $\ket{\bm{r}_1,\bm{r}_2} = \hat{b}^{\dag}_{\bm{r}_1}\hat{b}^{\dag}_{\bm{r}_2} \ket{n_{\rm A}=1}$, where the state $\ket{n_{\rm A}=1}$ represents the fully occupied A sublattice. 
Taking into account the translation invariance, a pair with the center of mass momentum $\bm{K}$ is described by $\ket{\phi(\bm{K},\bm{r})} = (1/\sqrt{N_{\rm s}})\sum_{\bm{R}_j} e^{i\bm{K}\cdot\bm{R}_j}\ket{\bm{R}_j, \bm{R}_j+\bm{r}}$, where $\bm{r}$ is the relative vector of two fermions and $N_{\rm s}$ is the number of unit cells~\cite{Crepel2021}.  
The lowest-energy state of two interacting fermions is obtained by solving the eigenvalue problem $\hat{H}_{\rm eff}\ket{\Phi(\bm{K})}=E_2(\bm{K})\ket{\Phi(\bm{K})}$, with $\ket{\Phi(\bm{K})}=\sum_{\bm{r}} C(\bm{r}) \ket{\phi(\bm{K},\bm{r})}$. 
Since $U_3$ does not work in the two-body problem, $t_{ij}$, $\lambda_{ij}$, and $V$ in Eq.~(\ref{eq:Heff}) determine the two-body state. 
The details of this two-body problem are given in Appendix~\ref{app:two_body}. 
Since the ground state in equilibrium is at $\bm{K}=\bm{0}$~\cite{Crepel2021}, we assume $\bm{K}=\bm{0}$ even in the effective model under CPL.  

\begin{figure}[t]
\centering
\includegraphics[width=0.9\columnwidth]{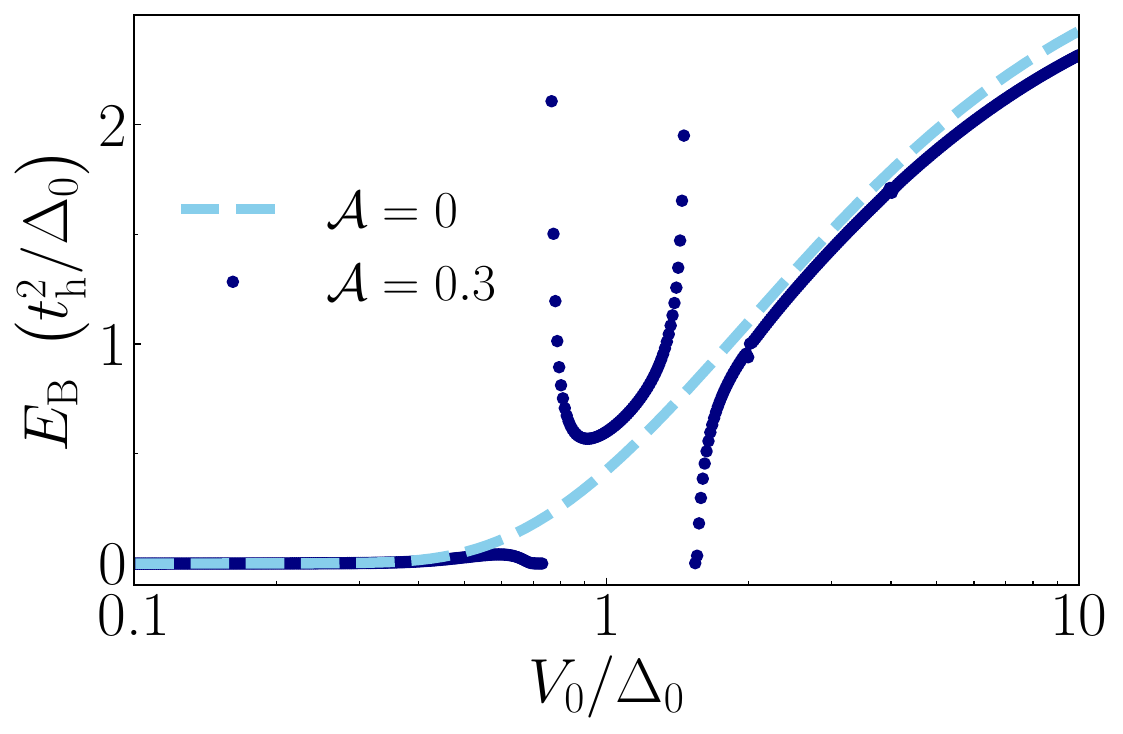}
\caption{Binding energy $E_{\rm{B}}$ for $\mathcal{A}=0$ and $\mathcal{A}=0.3$ at $\hbar\Omega/\Delta_0=2.5$.} 
\label{fig8}
\end{figure}

To see how CPL affects the pairing state, in Fig.~\ref{fig8}, we plot the binding energy defined as $E_{\rm{B}} = -[ E_{2}(\bm{K}=\bm{0}) - E_{1}(\bm{k}_{\rm min}) - E_{1}(-\bm{k}_{\rm min})  ]$, where $E_{2}(\bm{K}=\bm{0})$ is the lowest energy of the two-body state and $E_{1}(\bm{k}_{\rm min})$ is the single-particle energy at $\bm{k}_{\rm min}$ that gives the energy minimum of the noninteracting band made of $t_{ij}$.  
Note that $E_{1}(-\bm{k}_{\rm min})$ does not always coincide with the energy minimum [i.e., $E_{1}(-\bm{k}_{\rm min}) \ne E_{1}(\bm{k}_{\rm min})$] when $t_{ij}$ is complex. 
However, since ${\rm Im}[t_{ij}]$ is small except for the divergence point ($m \hbar\Omega=\Delta_0 + qV_0$), the difference between the one-body energies $E(\bm{k}_{\rm min})$ and $E(-\bm{k}_{\rm min})$ is small.  
Figure~\ref{fig8} shows $E_{\rm{B}}$ at $\hbar\Omega/\Delta_0 = 2.5$, where the attractive contribution to $V$ is induced around $V_{0}/\Delta_{0}=1$ [see Figs.~\ref{fig2}(c) and \ref{fig6}]. 
The binding energy can be nonzero due to the correlated hopping $\lambda_{ij}$ even though $V$ is zero or repulsive~\cite{Crepel2021}.  
As shown in Fig.~\ref{fig8}, $E_{\rm B}$ at $\mathcal{A}=0$ monotonically increases as $V_0$ increases. 
$E_{\rm B}$ under CPL ($\mathcal{A}=0.3$) shows a similar tendency to $E_{\rm B}$ at $\mathcal{A}=0$ but is enhanced at $0.75 < V_0/\Delta_0 < 1.5$ (i.e., $\Delta_0 + V_0 < \hbar \Omega < \Delta_0 + 2V_0$), which is associated with the enhancement of the attraction in $V$.  
We indeed find a signature of strong binding of doped fermions. 

\begin{figure}[t]
\centering
\includegraphics[width=0.9\columnwidth]{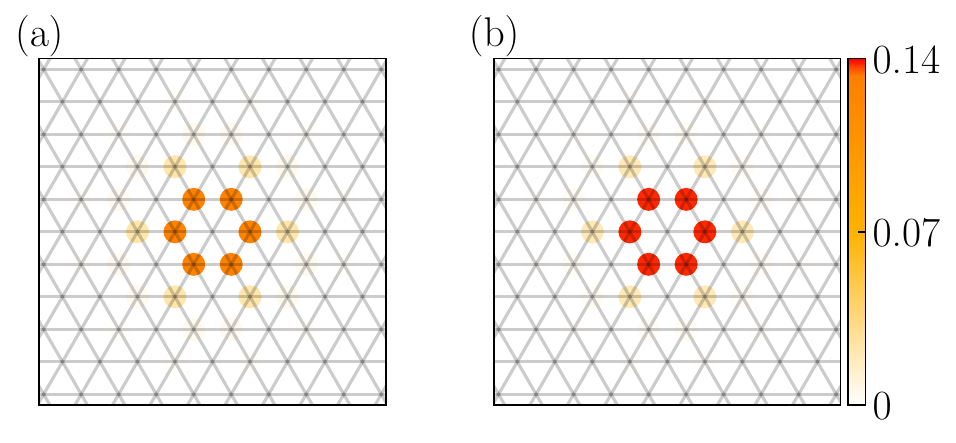}   
\\
\includegraphics[width=0.9\columnwidth]{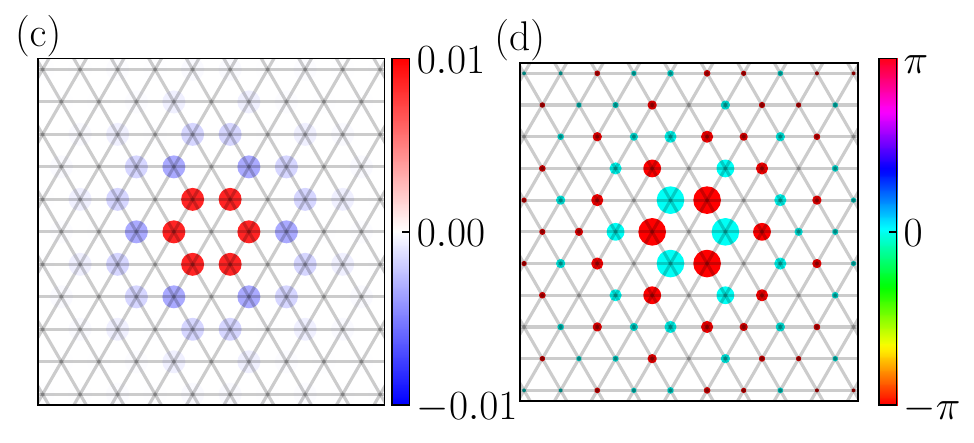}
\\
\includegraphics[width=0.8\columnwidth]{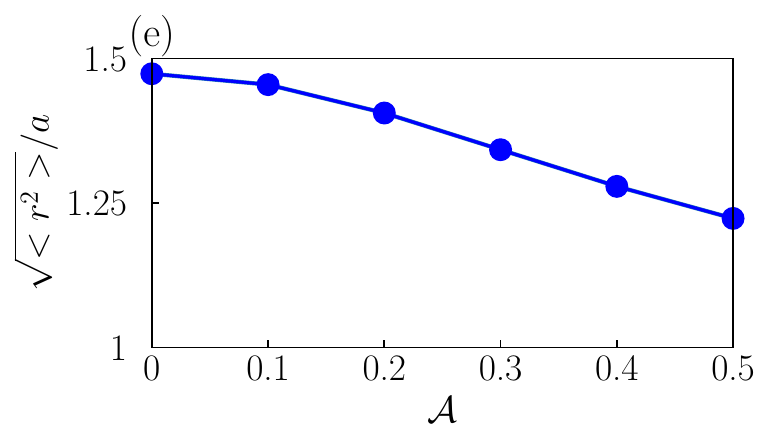}
\caption{Bound states at (a) $\mathcal{A}=0$ and (b) $\mathcal{A}=0.3$, where the densities  $|C(\bm{r})|^2$ are plotted. $V_0/\Delta_0=1.0$ and $\hbar\Omega/\Delta_0=2.5$ 
are used in the calculations.
(c) Density difference between $\mathcal{A}=0.3$ and $\mathcal{A}=0.0$. 
(d) $\arg C(\bm{r})$ at $\mathcal{A}=0.3$. The colors indicate $\arg C(\bm{r})$, and the size of each circle is proportional to $|C(\bm{r})|$. 
(e) $\sqrt{\braket{r^2}}$ as a function $\mathcal{A}$, where $a$ ($=\sqrt{3}d$) is the lattice constant.} 
\label{fig9}
\end{figure}

Furthermore, we present the real-space distribution of the two-body wave function in Fig.~\ref{fig9}.  
Figures~\ref{fig9}(a) and \ref{fig9}(b) show the densities given by $|C(\bm{r})|^{2}$ at $\mathcal{A}=0$ and $\mathcal{A}=0.3$, respectively. 
The difference between the densities with and without $\mathcal{A}$ is shown in Fig.~\ref{fig9}(c). 
At $V_0/\Delta_0=1$, $\mathcal{A}$ enhances the weights on the six nearest-neighbor sites but suppresses the weights away from them.
This indicates that CPL can strengthen the binding of fermions consistent with the enhancement of $E_{\rm B}$. 
To quantify the spatial extension of a pair, we compute $\braket{r^2} =\sum_{\bm{r}} r^2 |C(\bm{r})|^2$~\cite{Crepel2021}. 
As shown in Fig.~\ref{fig9}(e), $\sqrt{\braket{r^2}}$ decreases as $\mathcal{A}$ increases; i.e., the binding of two fermions becomes stronger in the effective model. 
These results suggest that CPL can support the formation of pairing states. 

To show the pairing symmetry, we plot $\arg C(\bm{r})$ at $\mathcal{A}=0.3$ in Fig.~\ref{fig9}(d). 
The wave function changes sign but is symmetric under threefold rotation.  
Hence, the bound state shown in Fig.~\ref{fig9}(d) has $f$-wave symmetry. 
This pairing symmetry is consistent with that in equilibrium without CPL ($\mathcal{A}=0$) shown in Ref.~\cite{Crepel2021}. 
Although the effective parameters $t_{ij}$ and $\lambda_{ij}$ are complex when $\mathcal{A} \ne 0$, the phase of $C(\bm{r})$ is $0$ or $\pi$; i.e., no noticeable imaginary contribution is observed in the two-body bound state shown in Fig.~\ref{fig9}.   
Since the weights at the next-nearest-neighboring B sites (at $|\bm{r}|=\sqrt{3}a=3d$) vanish, no signature of $(f + if')$-wave-like pairing~\cite{Zhou2023} is apparent. 
If the next-nearest-neighbor B-B hopping (which may appear in higher-order perturbation processes) is active, it may result in a more exotic pairing state~\cite{Kitamura2022}. 
In the parameter regime we addressed, ${\rm Im}[t_{ij}]$ and ${\rm Im}[\lambda_{ij}]$ are small, and the ground-state pairing properties strongly remain even in the effective model under CPL.


\section{Discussion}\label{sec.dis}

\begin{figure}[t]
\centering
\includegraphics[width=0.7\columnwidth]{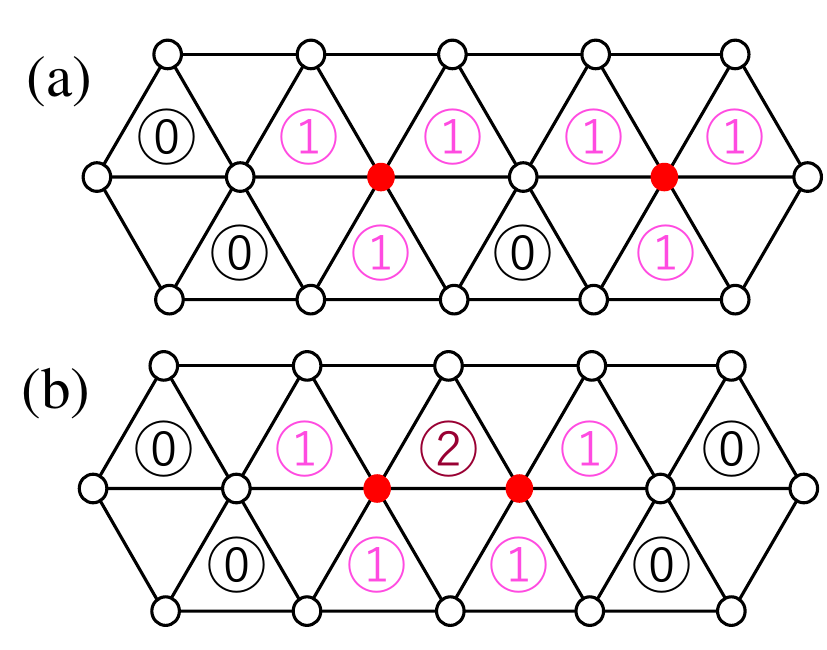}   
\caption{(a) Two separated fermions and (b) paired fermions.  
The red circles indicate fermions on the triangular B sublattice, and the number in the upper triangle represents $n^{\rm B}_{\triangle}$ $(=0, 1, 2)$. 
}
\label{fig10}
\end{figure}

Finally, we discuss the physical reason for the enhanced pairing interaction. 
Here, we consider the second-order energies that give the effective interactions (see Appendix \ref{app:der}). 
The perturbative energy on a triangle with $n^{\rm B}_{\triangle}$ $(=0,\;1,\;2)$ fermions is given by
\begin{align}
E(n^{\rm B}_{\triangle}) 
= - \sum^{\infty}_{m=-\infty} \frac{(3-n^{\rm B}_{\triangle})t_{\rm{h}}^2 J_{m}(\mathcal{A})^2}{\Delta_0 + (2-n^{\rm B}_{\triangle})V_{0} -m\hbar\Omega}, 
\end{align}
where $3-n^{\rm B}_{\triangle}$ in the numerator is the number of possible perturbation processes (i.e., the number of empty sites on a triangle). 
The pairing energy can be evaluated by comparing the energies of the states shown in Fig.~\ref{fig10}. 
From the state of two separated fermions in Fig.~\ref{fig10}(a) to the paired state in Fig.~\ref{fig10}(b), the number of the $n^{\rm B}_{\triangle}=1$ triangles decreases by $2$, while the numbers of the $n^{\rm B}_{\triangle}=0$ and $n^{\rm B}_{\triangle}=2$ triangles increase by $1$.  
Hence, the energy difference is given by $E(n^{\rm B}_{\triangle}=0) - 2E(n^{\rm B}_{\triangle}=1)+E(n^{\rm B}_{\triangle}=2)$, which is consistent with the two-body interaction $V$. 

The generated interaction under CPL (in the weak-$\mathcal{A}$ regime) is mainly caused by the sum of the $m= + 1$ and $m=-1$ terms, i.e., 
\begin{align}
E_1(n^{\rm B}_{\triangle})
= - \frac{2(3-n^{\rm B}_{\triangle})\left[\Delta_0 + (2-n^{\rm B}_{\triangle})V_{0}\right]t_{\rm{h}}^2 J_{1}(\mathcal{A})^2}{\left[\Delta_0 + (2-n^{\rm B}_{\triangle})V_{0}\right]^2 -(\hbar\Omega)^2}.  
\end{align}
As shown in Figs.~\ref{fig5} and ~\ref{fig6}, the attractive region of $V$ is generated at $\Delta_0+V_0 < \hbar\Omega < \Delta_0+2V_0$.
In this frequency range, $E_{1}(n^{\rm B}_{\triangle}=0) <0$, $E_{1}(n^{\rm B}_{\triangle}=1) >0$, and $E_{1}(n^{\rm B}_{\triangle}=2) >0$, suggesting that the $n^{\rm B}_{\triangle}=0$ triangle is energetically favorable, while the $n^{\rm B}_{\triangle}=1$ and 2 triangles are unfavorable. 
From Fig.~\ref{fig10}(a) to Fig.~\ref{fig10}(b), the pair formation creates one favorable $n^{\rm B}_{\triangle}=0$ triangle and one unfavorable $n^{\rm B}_{\triangle}=2$ triangle but annihilates two unfavorable $n^{\rm B}_{\triangle}=1$ triangles. 
Here, $E_1(n^{\rm B}_{\triangle}=0) + E_1(n^{\rm B}_{\triangle}=2)<0$ at $\Delta_0+V_0 < \hbar\Omega < \Delta_0+2V_0$; i.e., the energy loss of a $n^{\rm B}_{\triangle}=2$ triangle is smaller than the energy gain of a $n^{\rm B}_{\triangle}=0$ triangle.  
Therefore, the creation of one favorable $n^{\rm B}_{\triangle}=0$ triangle and the annihilation of two unfavorable $n^{\rm B}_{\triangle}=1$ triangles give rise to the attractive interaction between two fermions. 
The main cause of this phenomenon is the sign inversion of $E_{1}(n^{\rm B}_{\triangle}=1)$ due to $\hbar \Omega$.


\section{Summary and outlook}\label{sec.sum}

We derived an effective model for the correlated fermions on the honeycomb lattice under CPL. 
In particular, we estimated the effective hopping and many-body interactions for fermions doped on the band insulator.   
We found that the effective two-body interaction $V$ can obtain an attractive contribution with the optimal choice of the driving frequency $\Omega$.  
By solving the two-body problem, we demonstrated that the binding of two fermions, which has $f$-wave symmetry, becomes stronger following the enhancement of the attraction in $V$. 
Our effective model suggests that CPL can reinforce pairing states in strongly correlated systems.  

As shown in Sec.~\ref{sec.eff}, $V$ modified by CPL becomes attractive at $2\Delta_0 \lesssim \hbar \Omega \lesssim 3\Delta_0$ when $V_0 \sim \Delta_0$. 
For example, when $\Delta_0 \sim 2$~eV, $V$ is attractive at $\hbar \Omega \sim$ 5~eV. 
If CPL with $\hbar \Omega \sim$ 5~eV is applied to the system with the lattice distance $d \sim 2.5$~\AA, $\mathcal{A}=0.1$ corresponds to the electric field $E_0 = (\hbar \Omega/ed) \mathcal{A} \sim 20 $~MV/cm. 
In this case, a strong electric field is required to generate a strong attraction.   
A small $\Omega$ can suppress the magnitude of $E_0$, whereas $\Delta_0 \gg t_{\rm h}$ must be satisfied. 
This suggests that narrow band systems with small $t_{\rm h}$ are suitable for suppressing $E_0$. 
From this perspective, nearly flatband systems, e.g., in transition-metal dichalcogenide (TMD) heterobilayers~\cite{Crepel2023}, may be a candidate system (whereas the interplay between the spin-orbit coupling and CPL should be carefully considered in TMDs). 
As shown in Ref.~\cite{Crepel2022pnas}, the pair of spinless fermions we discussed corresponds to the spin-triplet pair in the spinful system, implying that the spin-triplet superconductivity has the potential to be enhanced by CPL. 
A study using a model with spin degrees of freedom will be an important issue in considering real materials.  

In the Floquet effective model under CPL, the complex hopping $t_{ij}$ breaks time-reversal symmetry in the spinless fermion system. 
In Fig.~\ref{fig9}, where the binding of fermions is strengthened, the ground-state pairing property strongly remains, and a noticeable imaginary contribution is not obtained in the two-body bound state.  
If ${\rm Im}[t_{ij}]$ (which may require large $\mathcal{A}$) is large, it can make the asymmetry between the $\bm{k}$ and $-\bm{k}$ points [i.e., $E_1(\bm{k}) \ne E_1(-\bm{k})$] larger and should affect pairing properties. 
Although we address only the two-body problem, sensitivity to the asymmetric band structure may also depend on band filling or the Fermi surface. 
If CPL strongly revises the pairing symmetry, the time evolution from the initial $f$-wave pairing state to a renewed pairing state is an intriguing issue. 
Time-dependent simulations in many-particle systems (e.g., using the exact method in finite-size systems~\cite{Claassen2017} and the Gutzwiller ansatz~\cite{Anan2024}) may provide fruitful insights into Floquet superconductivity.  
These demonstrations are interesting topics for future investigation.


\begin{acknowledgments}
This work was supported by Grants-in-Aid for Scientific Research from JSPS, KAKENHI Grants No.~JP20H01849, No.~JP24K06939, No.~JP24H00191 (T.K.), No.~JP22K04907 (K.K.), and No.~JP24K01333. 
\end{acknowledgments}


\appendix


\section{Derivation of the effective model} \label{app:der}

We conduct the two-step transformation using $e^{i\hat{\Lambda}(t)}= e^{i\hat{\Lambda}'(t)} e^{i\hat{S}(t)}$ to derive the effective Hamiltonian $\hat{H}_{\rm eff}$~\cite{Kitamura2022}. 
$\hat{S}(t)$ is used for the strong-coupling expansion to map the original Hamiltonian $\hat{H}(t)$ into the effective interacting Hamiltonian $\hat{H}'(t)$ defined on the triangular network of the B sublattice. 
Then, $\hat{\Lambda}'(t)$ is used to make the effective static Hamiltonian, i.e., $\hat{H}'(t) \rightarrow \hat{H}_{\rm eff}$~\cite{Kitamura2022}. 

First, we conduct the strong-coupling expansion based on the transformation $\hat{H}'(t) = e^{i\hat{S}(t)} [ \hat{H}(t) -i \hbar \partial_t] e^{-i\hat{S}(t)}$. 
To derive the effective model at $\Delta_0 \gg t_{\rm h}$, we decompose the hopping term of Eq.~(\ref{eq:ham_A}) considering the energy levels of the intermediate states in the perturbation processes. 
In the spinless model, there are three types of excited states, whose energies depend on the occupancy of fermions on a triangle (three B sites) around an A site. 
The energy changes associated with the hopping from A to B sites (without the external field) are $\Delta_0$ when two of three sites are occupied, $\Delta_0+V_0$ when one of three sites is occupied, and $\Delta_0 + 2V_0$ when three sites are unoccupied. 
These three conditions can be classified by the projection operators defined as  
\begin{align}
&\hat{P}^{\rm B}_{0;kl} =  \hat{n}_{\bm{r}+\bm{d}_{k}}^{\rm B}\hat{n}_{\bm{r}+\bm{d}_{l}}^{\rm B}, 
\\
&\hat{P}^{\rm B}_{1;kl} = \hat{n}_{\bm{r}+\bm{d}_{k}}^{\rm B}(1-\hat{n}_{\bm{r}+\bm{d}_{l}}^{\rm B}) + (1-\hat{n}_{\bm{r}+\bm{d}_{k}}^{\rm B})\hat{n}_{\bm{r}+\bm{d}_{l}}^{\rm B} , 
\\
&\hat{P}^{\rm B}_{2;kl} = (1-\hat{n}_{\bm{r}+\bm{d}_{k}}^{\rm B})(1-\hat{n}_{\bm{r}+\bm{d}_{l}}^{\rm B}).  
\end{align} 
The subscript $q$ in $\hat{P}^{\rm B}_{q;kl}$ reflects the energy change of $\Delta_0 + q V_0$, where $q=2-n^{\rm B}_{\triangle}$ with respect to the number of fermions on a triangle $n^{\rm B}_{\triangle}$.  

Using these projection operators and the Jacobi-Anger expansion, the hopping under CPL is written as 
\begin{align}
&\hat{T}_{\rm h} (t) = \sum_{m=-\infty}^{\infty} \sum_{q=0}^{2} \left( \hat{T}_{+1,q,m} + \hat{T}_{-1,-q,m} \right)e^{-im\Omega t}, 
\end{align}
with 
\begin{align}
\hat{T}_{+1,q,m} =& -t_{\rm h} \sum_{\bm{r} \in \rm A}\sideset{}{'}{\sum}_{(j,k,l)}(-i)^{m}J_{m}(\mathcal{A})e^{-im\alpha_{j}}\hat{b}_{\bm{r}+\bm{d}_{j}}^{\dagger}\hat{a}_{\bm{r}}\hat{P}^{\rm B}_{q;kl}, 
\end{align}
where $\hat{T}_{-1,-q,m}$ ($=\hat{T}^{\dag}_{+1,q,-m}$) indicates the hopping from B to A sites and the primed sum means $\displaystyle\sideset{}{'}{\sum}_{(j,k,l)} F_{jkl} = F_{123} + F_{231} + F_{312}$. 
The hopping $\hat{T}_{\pm 1,\pm q, m}$ satisfies  
\begin{align}
&\left[\Delta_{0}\sum_{\bm{r} \in \rm B} \hat{n}_{\bm{r}}^{\rm B},\; \hat{T}_{\pm 1,\pm q, m}\right]
=\pm \Delta_{0}\hat{T}_{\pm 1,\pm q,m} ,
\\
&\left[V_{0}\sum_{\bm{r}\in \rm A}\sum_{j=1,2,3} \hat{n}_{\bm{r}}^{\rm A}\hat{n}_{\bm{r}+\bm{d}_{j}}^{\rm B},\; \hat{T}_{\pm 1,\pm q, m}\right]
=\pm qV_{0}\hat{T}_{\pm 1,\pm q,m}. 
\end{align}
Hence, 
\begin{align}
\left[\hat{H}_0,\; \hat{T}_{\pm 1,\pm q, m}\right]
=\pm \left( \Delta_{0} + q V_0 \right) \hat{T}_{\pm 1,\pm q,m}. 
\label{eq:comm_H0}
\end{align}
This commutation relation indicates that the hopping $\hat{T}_{\pm 1,\pm q,0}$ changes the energy $\pm ( \Delta_0 + q V_0)$.  

A series expansion $\hat{S}(t) = \hat{S}^{(1)}(t) + \hat{S}^{(2)}(t) \cdots$ [where $\hat{S}^{(n)}(t) \propto t_{\rm h}^n$] gives~\cite{Kitamura2016,Eckstein2017} 
\begin{align}
\hat{H}'(t) 
&= \hat{H}_0 + \hat{T}_{\rm h}(t) 
+ \left[ i\hat{S}^{(1)}(t), \hat{H}_0 \right] 
 - \hbar \frac{\partial \hat{S}^{(1)}(t)}{\partial t}
\notag \\ 
&+ \left[ i\hat{S}^{(1)}(t), \hat{T}_{\rm h}(t)\right] 
+ \left[ i\hat{S}^{(2)}(t), \hat{H}_0 \right] 
 - \hbar \frac{\partial \hat{S}^{(2)}(t)}{\partial t}
\notag \\
&+ \frac{1}{2}\left[ i\hat{S}^{(1)}(t), \left[ i\hat{S}^{(1)} , \hat{H}_0 \right] 
-\hbar \frac{\partial \hat{S}^{(1)}(t)}{\partial t}
\right]  + \cdots. 
\end{align}
The AB hopping at the first order is excluded by setting $\hat{S}^{(1)}(t)$ to satisfy 
\begin{align}
\hat{T}_{\rm h}(t) 
+ \left[ i\hat{S}^{(1)}(t), \hat{H}_0 \right] 
 - \hbar \frac{\partial \hat{S}^{(1)}(t)}{\partial t} = 0. 
\label{eq:perturb_1st}
\end{align}
Using Eq.~(\ref{eq:comm_H0}), we find that 
\begin{align}
i \hat{S}^{(1)}(t) 
&= \sum_{m=-\infty}^{\infty} \sum_{q=0}^{2}
\frac{\hat{T}_{+1,q,m}}{\Delta_{0} + qV_{0} - m \hbar \Omega} e^{-im\Omega t}
\notag \\
& -  \sum_{m=-\infty}^{\infty} \sum_{q=0}^{2}
\frac{\hat{T}_{-1,-q,m}}{\Delta_{0} + qV_{0} + m \hbar \Omega} 
e^{-im\Omega t}
\end{align}
can satisfy Eq.~(\ref{eq:perturb_1st}). 
Then, the transformed Hamiltonian becomes 
\begin{align}
\hat{H}'(t) 
= \hat{H}_0 & + \frac{1}{2}\left[ i\hat{S}^{(1)}(t), \hat{T}_{\rm h}(t)\right] 
\notag \\
&+ \left[ i\hat{S}^{(2)}(t), \hat{H}_0 \right] 
 - \hbar \frac{\partial \hat{S}^{(2)}(t)}{\partial t}
+  \cdots. 
\label{eq:H_2nd_S2}
\end{align}
Here, $[ i\hat{S}^{(1)}(t), \hat{T}_{\rm h}(t)]$ in the second term includes $[ \hat{T}_{\pm 1,\pm q,m}, \hat{T}_{\mp1,\mp q,n} ]$ and $[ \hat{T}_{\pm 1,\pm q,m}, \hat{T}_{\pm1,\pm q,n} ]$ terms. 
To address the subspace, in which fermions fully occupy the A sublattice, we exclude $[ \hat{T}_{\pm 1,\pm q,m}, \hat{T}_{\pm1,\pm q,n} ]$ terms by an optimal choice of $\hat{S}^{(2)}(t)$ in the third and fourth terms of Eq.~(\ref{eq:H_2nd_S2}). 
Then, we obtain the effective Hamiltonian up to the second order, 
\begin{align}
\hat{H}'(t) 
\simeq \hat{H}_0 
+ \frac{1}{2}\sum_{m,n} \sum_{q=0}^{2}
 \frac{ \left[ \hat{T}_{+1,q,m}, \hat{T}_{-1,-q,n-m} \right]}{\Delta_{0} + qV_{0} - m \hbar \Omega} e^{-in\Omega t}
\notag \\
- \frac{1}{2} \sum_{m,n} \sum_{q=0}^{2}
\frac{\left[ \hat{T}_{-1,-q,m},\hat{T}_{+1,q,n-m}  \right]}{\Delta_{0} + qV_{0} + m \hbar \Omega}  e^{-i n \Omega t}. 
\label{eq:H_2nd_n}
\end{align}

$\hat{H}'(t)$ is a time-periodic Hamiltonian due to $e^{-in\Omega t}$, and its Floquet states are characterized by $n\Omega$. 
To derive a static Hamiltonian, we conduct the transformation 
$\hat{H}_{\rm eff} = e^{i\hat{\Lambda}'(t)} [ \hat{H}'(t) -i \hbar \partial_t] e^{-i\hat{\Lambda}'(t)}$~\cite{Kitamura2022}. 
When the energy interval of $\Omega$ is much larger than the energy scale of the second-order term, which is on the order of $t_{\rm h}^2/(\Delta_0 + qV_0 - m \hbar \Omega)$, the mapping to the $n=0$ subspace without the high-frequency expansion terms may give a reasonable effective Hamiltonian. 
Neglecting the higher-order perturbation terms, the effective static Hamiltonian up to the second order is given by 
\begin{align}
\hat{H}_{\rm eff} 
&=  \hat{H}_0 
+ \sum_{m=-\infty}^{\infty} \sum_{q=0}^{2}
 \frac{ \left[ \hat{T}_{+1,q,m}, \hat{T}_{-1,-q,-m} \right]}{\Delta_{0} + qV_{0} - m \hbar \Omega} .
\end{align} 
Because we consider the case where the A sublattice is fully occupied, $\hat{H}_0$ gives the energy $N_{\rm B} (\Delta_0 + 3V_0)$ when $N_{\rm B}$ fermions are doped. 
The $\hat{T}_{+1,q,m}\hat{T}_{-1,-q,-m}$ term vanishes when all A sites are occupied. 
Hence, the effective Hamiltonian at the second order is given by 
\begin{align}
\hat{H}^{(2)}_{\rm eff} 
=  - \sum_{m=-\infty}^{\infty} \sum_{q=0}^{2}
 \frac{\hat{T}_{-1,-q,-m} \hat{T}_{+1,q,m} }{\Delta_{0} + qV_{0} - m \hbar \Omega} .
\end{align}

Assuming that all A sites are occupied, this effective Hamiltonian is written as 
\begin{align}
\hat{H}^{(2)}_{\rm eff} 
=  - \sum_{m=-\infty}^{\infty} \sum_{q=0}^{2}
\sum_{\bm{r} \in \rm A}\sideset{}{'}{\sum}_{(j,k,l)}\sideset{}{'}{\sum}_{(j',k',l')}  
t_{q,m} e^{im(\alpha_{j}-\alpha_{j'})}  
\notag \\
\times \hat{b}_{\bm{r}+\bm{d}_{j}} \hat{P}^{\rm B}_{q;kl}
\hat{P}^{\rm B}_{q;k'l'} \hat{b}_{\bm{r}+\bm{d}_{j'}}^{\dagger}
\end{align}
with 
\begin{align}
t_{q,m} = \frac{t^2_{\rm h} J_{m}(\mathcal{A})^2 }{\Delta_{0} + qV_{0} - m \hbar \Omega} . 
\end{align}
Depending on the index $q$, 
\begin{align}
&\sideset{}{'}{\sum}_{(j,k,l)}\sideset{}{'}{\sum}_{(j',k',l')} e^{im(\alpha_{j}-\alpha_{j'})}  
\hat{b}_{\bm{r}+\bm{d}_{j}} \hat{P}^{\rm B}_{0;kl} \hat{P}^{\rm B}_{0;k'l'} \hat{b}_{\bm{r}+\bm{d}_{j'}}^{\dagger}
\notag \\
& = \sideset{}{'}{\sum}_{(j,k,l)} ( 1-\hat{n}_{\bm{r}+\bm{d}_{j}}^{\rm B} ) \hat{n}_{\bm{r}+\bm{d}_{k}}^{\rm B}\hat{n}_{\bm{r}+\bm{d}_{l}}^{\rm B}
\notag \\
& - \sideset{}{'}{\sum}_{(j,k,l)} \left[ e^{im(\alpha_{j}-\alpha_{k})}  
\hat{b}_{\bm{r}+\bm{d}_{k}}^{\dagger}\hat{b}_{\bm{r}+\bm{d}_{j}}\hat{n}^{\rm B}_{\bm{r}+\bm{d}_l}
+ {\rm H.c.}
\right], 
\notag \\
&\sideset{}{'}{\sum}_{(j,k,l)}\sideset{}{'}{\sum}_{(j',k',l')} e^{im(\alpha_{j}-\alpha_{j'})}  
\hat{b}_{\bm{r}+\bm{d}_{j}} \hat{P}^{\rm B}_{1;kl} \hat{P}^{\rm B}_{1;k'l'} \hat{b}_{\bm{r}+\bm{d}_{j'}}^{\dagger}
\notag \\
& = \sideset{}{'}{\sum}_{(j,k,l)} 
2 ( 1-\hat{n}_{\bm{r}+\bm{d}_{j}}^{\rm B} ) 
(1-\hat{n}_{\bm{r}+\bm{d}_{k}}^{\rm B})\hat{n}_{\bm{r}+\bm{d}_{l}}^{\rm B}
\notag \\
& - \sideset{}{'}{\sum}_{(j,k,l)} \left[ e^{im(\alpha_{j}-\alpha_{k})}  
\hat{b}_{\bm{r}+\bm{d}_{k}}^{\dagger} \hat{b}_{\bm{r}+\bm{d}_{j}}(1 - \hat{n}^{\rm B}_{\bm{r}+\bm{d}_l})
+ {\rm H.c.}
\right], 
\notag 
\end{align}
and 
\begin{align}
&\sideset{}{'}{\sum}_{(j,k,l)}\sideset{}{'}{\sum}_{(j',k',l')} e^{im(\alpha_{j}-\alpha_{j'})}  
\hat{b}_{\bm{r}+\bm{d}_{j}} \hat{P}^{\rm B}_{2;kl} \hat{P}^{\rm B}_{2;k'l'} \hat{b}_{\bm{r}+\bm{d}_{j'}}^{\dagger}
\notag \\
& = \sideset{}{'}{\sum}_{(j,k,l)} ( 1-\hat{n}_{\bm{r}+\bm{d}_{j}}^{\rm B} ) (1-\hat{n}_{\bm{r}+\bm{d}_{k}}^{\rm B} ) (1-\hat{n}_{\bm{r}+\bm{d}_{l}}^{\rm B} ) . 
\notag 
\end{align}
\begin{widetext}
Using the above equations, we obtain 
\begin{align}
\hat{H}^{(2)}_{\rm eff} 
&= \sum_{m=-\infty}^{\infty} 
3\left( t_{0,m} - 2 t_{1,m} + t_{2,m} \right) 
\sum_{\bm{r} \in \rm A} \hat{n}_{\bm{r}+\bm{d}_{1}}^{\rm B}\hat{n}_{\bm{r}+\bm{d}_{2}}^{\rm B}\hat{n}_{\bm{r}+\bm{d}_{3}}^{\rm B}
\notag \\
& + \sum_{m=-\infty}^{\infty} 
\left( - t_{0,m} + 4t_{1,m} - 3t_{2,m} \right) 
\sum_{\bm{r} \in \rm A}\left( 
\hat{n}_{\bm{r}+\bm{d}_{1}}^{\rm B} \hat{n}_{\bm{r}+\bm{d}_{2}}^{\rm B}
+\hat{n}_{\bm{r}+\bm{d}_{2}}^{\rm B} \hat{n}_{\bm{r}+\bm{d}_{3}}^{\rm B}
+\hat{n}_{\bm{r}+\bm{d}_{3}}^{\rm B} \hat{n}_{\bm{r}+\bm{d}_{1}}^{\rm B}
\right)
\notag \\
& + \sum_{m=-\infty}^{\infty} 
\left( - 2t_{1,m} + 3t_{2,m} \right) 
\sum_{\bm{r} \in \rm A} \left( 
\hat{n}_{\bm{r}+\bm{d}_{1}}^{\rm B} + \hat{n}_{\bm{r}+\bm{d}_{2}}^{\rm B} + \hat{n}_{\bm{r}+\bm{d}_{3}}^{\rm B}
\right)
- 3N_{\rm A}\sum_{m=-\infty}^{\infty} t_{2,m} 
\notag \\
& + \sum_{m=-\infty}^{\infty} 
t_{1,m}
\sum_{\bm{r} \in \rm A} \left[e^{i\frac{2m\pi}{3}}\left(
\hat{b}^{\dag}_{\bm{r}+\bm{d}_2}\hat{b}_{\bm{r}+\bm{d}_1} 
+\hat{b}^{\dag}_{\bm{r}+\bm{d}_3}\hat{b}_{\bm{r}+\bm{d}_2}
+\hat{b}^{\dag}_{\bm{r}+\bm{d}_1}\hat{b}_{\bm{r}+\bm{d}_3}
\right)
+{\rm H.c.}
\right]
\notag \\
& + \sum_{m=-\infty}^{\infty} 
\left( t_{0,m}  - t_{1,m} \right)
\sum_{\bm{r} \in \rm A} \left[e^{i\frac{2m\pi}{3}}\left(
\hat{b}^{\dag}_{\bm{r}+\bm{d}_2}\hat{n}_{\bm{r}+\bm{d}_3}^{\rm B}\hat{b}_{\bm{r}+\bm{d}_1} 
+\hat{b}^{\dag}_{\bm{r}+\bm{d}_3}\hat{n}_{\bm{r}+\bm{d}_1}^{\rm B}\hat{b}_{\bm{r}+\bm{d}_2}
+\hat{b}^{\dag}_{\bm{r}+\bm{d}_1}\hat{n}_{\bm{r}+\bm{d}_2}^{\rm B}\hat{b}_{\bm{r}+\bm{d}_3}
\right)
+{\rm H.c.}
\right].
\end{align}
\end{widetext}
In this Hamiltonian, three-site interaction is given by
\begin{align}
U_3 = \sum_{m=-\infty}^{\infty} 3\left( t_{0,m} - 2 t_{1,m} + t_{2,m} \right), 
\end{align}
and two-site interaction is given by
\begin{align}
V = \sum_{m=-\infty}^{\infty} 
\left( - t_{0,m} + 4t_{1,m} - 3t_{2,m} \right) . 
\end{align}
The hopping of a fermion on the B sublattice is given by
\begin{align}
t_{ij} = \sum_{m=-\infty}^{\infty} t_{1,m} e^{i \nu_{ij} \frac{2m\pi}{3}}, 
\end{align}
and the correlated hopping is given by
\begin{align}
\lambda_{ij} = \sum_{m=-\infty}^{\infty} \left( t_{0,m}  - t_{1,m} \right) e^{i\nu_{ij}\frac{2m\pi}{3}}, 
\end{align}
where 
\begin{align}
\nu_{ij} = \bm{e}_z \cdot \frac{\bm{d}_i \times \bm{d}_j}{|\bm{d}_i \times \bm{d}_j|}. 
\end{align}

$\hat{H}^{(2)}_{\rm eff}$ includes the on-site energy term $\hat{H}^{(2)}_{\rm eff;0} =  - N_{\rm A}\Delta^{(2)}_{\rm A} + \Delta^{(2)}_{\rm B} \sum_{\bm{r} \in {\rm B}} \hat{n}^{\rm B}_{\bm{r}}$, with $\Delta^{(2)}_{\rm A} = \sum_{m} 3 t_{2,m}$ and $\Delta^{(2)}_{\rm B} = \sum_{m} 3\left( - 2t_{1,m} + 3t_{2,m} \right) $. 
This on-site term shifts the total energy. 
Since the A sites are fully occupied, the number of fermions on the A sublattice $N_{\rm A}$ is the same as the number of unit cells $N_{\rm s}$, and $N_{\rm B}$ fermions are doped on the B sublattice. 
Including the unperturbed term, $\hat{H}_0 + \hat{H}^{(2)}_{\rm eff;0}$ gives the energy $E_{\rm eff;0} = N_{\rm B} ( \Delta_0 + 3V_0 + \Delta^{(2)}_{\rm B} ) - N_{\rm A}\Delta^{(2)}_{\rm A}$ for $N_{\rm A} + N_{\rm B}$ fermions. 
We omit $\hat{H}_0 + \hat{H}^{(2)}_{\rm eff;0}$, which gives the constant $E_{\rm eff;0}$ from Eq.~(\ref{eq:Heff}).


\section{Two-body problem} \label{app:two_body}

Here, we present supplemental equations for the two-body problem.  
Assuming translation invariance, we introduce the two-body state 
\begin{align}
\ket{\phi(\bm{K},\bm{r})} = \frac{1}{\sqrt{N_{\rm s}}}\sum_{\bm{R}_j} e^{i\bm{K}\cdot \bm{R}_j} \hat{b}^{\dag}_{\bm{R}_j} \hat{b}^{\dag}_{\bm{R}_j+\bm{r}} \ket{n_{\rm A}=1}, 
\end{align}
where $\ket{n_{\rm A}=1}$ represents the fully occupied A sublattice. 
In the triangular B sublattice, the translation vectors are given by 
$\bm{a}_1 = \sqrt{3} d (1/2,\sqrt{3}/2)$, $\bm{a}_2 = \sqrt{3} d (1/2,-\sqrt{3}/2)$, and $\bm{a}_3 = -\bm{a}_1 - \bm{a}_2 =\sqrt{3}d (-1,0)$. 
In this appendix, the hopping parameters in $\hat{H}_{\rm eff}$ are denoted as $t_{ij} = t_b$ and $\lambda_{ij} = \lambda$ for $\bm{R}_j \rightarrow \bm{R}_j + \bm{a}_l$ (while $t_{ij} = t^*_b$ and 
$\lambda_{ij} = \lambda^*$
for $\bm{R}_j + \bm{a}_l \rightarrow \bm{R}_j$). 
To evaluate a bound state in the effective model, we consider $\hat{H}_{\rm{eff}}\ket{\phi(\bm{K},\bm{r})}$. 
The three-body interaction $U_3$ does not work in the present two-fermion system. 
Then, for the effective Hamiltonian  
\begin{align}
\hat{H}_{\rm{eff}}
&= \sum_{\bm{R}_j} \sum_{l=1,2,3} 
\left( 
t_{b}\hat{b}_{\bm{R}_j+\bm{a}_{l}}^{\dag}\hat{b}_{\bm{R}_j} + {\rm H.c.} 
\right)
\nonumber\\
&+ \sum_{\bm{R}_j} \sum_{l=1,2,3}  \left( 
\lambda{\hat{b}_{\bm{R}_j-\bm{a}_{l-1}}^{\dag}\hat{n}_{\bm{R}_j}\hat{b}_{\bm{R}_j+\bm{a}_{l}}} + {\rm H.c.} 
\right)
\nonumber\\
&+ V \sum_{\bm{R}_j} \sum_{l=1,2,3}\hat{{n}}_{\bm{R}_j}\hat{n}_{\bm{R}_j+\bm{a}_{l}} , 
\label{eq:Heff_two}
\end{align}
we derive  
\begin{align}
&\hat{H}_{\rm{eff}}\ket{\phi(\bm{K},\bm{r})}
=V\sum_{\epsilon= \pm}\sum_{l=1,2,3} \delta_{\bm{r},\epsilon \bm{a}_l} \ket{\phi(\bm{K},\bm{r})}
\nonumber\\
+&\sum_{l}\left[ 
\tau_{\bm{K},l} \ket{\phi(\bm{K},\bm{r}+\bm{a}_{l})} 
+ \tau_{\bm{K},l}^{*}\ket{\phi(\bm{K},\bm{r}-\bm{a}_{l})}  
\right]
\nonumber \\
+& \lambda \sum_{l} \delta_{\bm{r},\bm{a}_{l}} \ket{\phi(\bm{K},-\bm{a}_{l-1})} 
+ \lambda^{*} \sum_{l} \delta_{\bm{r},-\bm{a}_{l}} \ket{\phi(\bm{K},\bm{a}_{l+1})}
\nonumber \\
+ & \lambda \sum_{l} 
\delta_{\bm{r},-\bm{a}_{l}} e^{-i\bm{K} \cdot \bm{a}_{l+1}}\ket{\phi(\bm{K},\bm{a}_{l-1})}
\nonumber \\
+& \lambda^* \sum_{l} 
\delta_{\bm{r},\bm{a}_{l}} e^{i\bm{K} \cdot \bm{a}_{l-1}}\ket{\phi(\bm{K},-\bm{a}_{l+1})},
\label{eq:Heff_mat}
\end{align}
with $\tau_{\bm{K},l} = t_{b} + t_{b}^{*} e^{i \bm{K}\cdot \bm{a}_{l}}$. 
Here, $\bm{a}_{l-1}=\bm{a}_3$, with $l=1$, and $\bm{a}_{l+1}=\bm{a}_1$, with $l=3$. 
An eigenstate of the matrix composed of $\braket{\phi(\bm{K},\bm{r}') | \hat{H}_{\rm{eff}} | \phi(\bm{K},\bm{r})}$ gives a two-body bound state. 

We numerically solve this eigenvalue problem at $\bm{K}=\bm{0}$. 
We take 10980 lattice points for $\bm{r}$ and numerically diagonalize the matrix to obtain the eigenstates.  
Although extrapolations to the thermodynamic limit are necessary to obtain the exact results in the weak-coupling regime at $V_0/\Delta_0 < 0.25$~\cite{Crepel2021}, 
we present the results for the finite $\bm{r}$ points. 
The system size used in our calculation is sufficiently large to describe the bound states in the intermediate-coupling ($V_0 /\Delta_0 \sim 1$) and strong-coupling ($V_0/\Delta_0 \gg 1$) regimes, where the size of a pair is small and the system-size dependence is negligible.


\bibliography{reference}

\begin{thebibliography}{57}%
\makeatletter
\providecommand \@ifxundefined [1]{%
 \@ifx{#1\undefined}
}%
\providecommand \@ifnum [1]{%
 \ifnum #1\expandafter \@firstoftwo
 \else \expandafter \@secondoftwo
 \fi
}%
\providecommand \@ifx [1]{%
 \ifx #1\expandafter \@firstoftwo
 \else \expandafter \@secondoftwo
 \fi
}%
\providecommand \natexlab [1]{#1}%
\providecommand \enquote  [1]{``#1''}%
\providecommand \bibnamefont  [1]{#1}%
\providecommand \bibfnamefont [1]{#1}%
\providecommand \citenamefont [1]{#1}%
\providecommand \href@noop [0]{\@secondoftwo}%
\providecommand \href [0]{\begingroup \@sanitize@url \@href}%
\providecommand \@href[1]{\@@startlink{#1}\@@href}%
\providecommand \@@href[1]{\endgroup#1\@@endlink}%
\providecommand \@sanitize@url [0]{\catcode `\\12\catcode `\$12\catcode
  `\&12\catcode `\#12\catcode `\^12\catcode `\_12\catcode `\%12\relax}%
\providecommand \@@startlink[1]{}%
\providecommand \@@endlink[0]{}%
\providecommand \url  [0]{\begingroup\@sanitize@url \@url }%
\providecommand \@url [1]{\endgroup\@href {#1}{\urlprefix }}%
\providecommand \urlprefix  [0]{URL }%
\providecommand \Eprint [0]{\href }%
\providecommand \doibase [0]{https://doi.org/}%
\providecommand \selectlanguage [0]{\@gobble}%
\providecommand \bibinfo  [0]{\@secondoftwo}%
\providecommand \bibfield  [0]{\@secondoftwo}%
\providecommand \translation [1]{[#1]}%
\providecommand \BibitemOpen [0]{}%
\providecommand \bibitemStop [0]{}%
\providecommand \bibitemNoStop [0]{.\EOS\space}%
\providecommand \EOS [0]{\spacefactor3000\relax}%
\providecommand \BibitemShut  [1]{\csname bibitem#1\endcsname}%
\let\auto@bib@innerbib\@empty
\bibitem [{\citenamefont {Basov}\ \emph {et~al.}(2017)\citenamefont {Basov},
  \citenamefont {Averitt},\ and\ \citenamefont {Hsieh}}]{Basov2017}%
  \BibitemOpen
  \bibfield  {author} {\bibinfo {author} {\bibfnamefont {D.~N.}\ \bibnamefont
  {Basov}}, \bibinfo {author} {\bibfnamefont {R.~D.}\ \bibnamefont {Averitt}},\
  and\ \bibinfo {author} {\bibfnamefont {D.}~\bibnamefont {Hsieh}},\ }\bibfield
   {title} {\bibinfo {title} {Towards properties on demand in quantum
  materials},\ }\href {https://doi.org/https://doi.org/10.1038/nmat5017}
  {\bibfield  {journal} {\bibinfo  {journal} {Nat. Matter.}\ }\textbf {\bibinfo
  {volume} {16}},\ \bibinfo {pages} {1077} (\bibinfo {year}
  {2017})}\BibitemShut {NoStop}%
\bibitem [{\citenamefont {Ishihara}(2019)}]{Ishihara2019}%
  \BibitemOpen
  \bibfield  {author} {\bibinfo {author} {\bibfnamefont {S.}~\bibnamefont
  {Ishihara}},\ }\bibfield  {title} {\bibinfo {title} {Photoinduced {U}ltrafast
  {P}henomena in {C}orrelated {E}lectron {M}agnets},\ }\href
  {https://doi.org/10.7566/JPSJ.88.072001} {\bibfield  {journal} {\bibinfo
  {journal} {J. Phys. Soc. Jpn.}\ }\textbf {\bibinfo {volume} {88}},\ \bibinfo
  {pages} {072001} (\bibinfo {year} {2019})}\BibitemShut {NoStop}%
\bibitem [{\citenamefont {de~la Torre}\ \emph {et~al.}(2021)\citenamefont
  {de~la Torre}, \citenamefont {Kennes}, \citenamefont {Claassen},
  \citenamefont {Gerber}, \citenamefont {McIver},\ and\ \citenamefont
  {Sentef}}]{delaTorre2021}%
  \BibitemOpen
  \bibfield  {author} {\bibinfo {author} {\bibfnamefont {A.}~\bibnamefont
  {de~la Torre}}, \bibinfo {author} {\bibfnamefont {D.~M.}\ \bibnamefont
  {Kennes}}, \bibinfo {author} {\bibfnamefont {M.}~\bibnamefont {Claassen}},
  \bibinfo {author} {\bibfnamefont {S.}~\bibnamefont {Gerber}}, \bibinfo
  {author} {\bibfnamefont {J.~W.}\ \bibnamefont {McIver}},\ and\ \bibinfo
  {author} {\bibfnamefont {M.~A.}\ \bibnamefont {Sentef}},\ }\bibfield  {title}
  {\bibinfo {title} {Colloquium: Nonthermal pathways to ultrafast control in
  quantum materials},\ }\href {https://doi.org/10.1103/RevModPhys.93.041002}
  {\bibfield  {journal} {\bibinfo  {journal} {Rev. Mod. Phys.}\ }\textbf
  {\bibinfo {volume} {93}},\ \bibinfo {pages} {041002} (\bibinfo {year}
  {2021})}\BibitemShut {NoStop}%
\bibitem [{\citenamefont {Bloch}\ \emph {et~al.}(2022)\citenamefont {Bloch},
  \citenamefont {Cavalleri}, \citenamefont {Galitski}, \citenamefont {Hafezi},\
  and\ \citenamefont {Rubio}}]{Bloch2022}%
  \BibitemOpen
  \bibfield  {author} {\bibinfo {author} {\bibfnamefont {J.}~\bibnamefont
  {Bloch}}, \bibinfo {author} {\bibfnamefont {A.}~\bibnamefont {Cavalleri}},
  \bibinfo {author} {\bibfnamefont {V.}~\bibnamefont {Galitski}}, \bibinfo
  {author} {\bibfnamefont {M.}~\bibnamefont {Hafezi}},\ and\ \bibinfo {author}
  {\bibfnamefont {A.}~\bibnamefont {Rubio}},\ }\bibfield  {title} {\bibinfo
  {title} {Strongly correlated electron--photon systems},\ }\href
  {https://doi.org/10.1038/s41586-022-04726-w} {\bibfield  {journal} {\bibinfo
  {journal} {Nature}\ }\textbf {\bibinfo {volume} {606}},\ \bibinfo {pages}
  {41} (\bibinfo {year} {2022})}\BibitemShut {NoStop}%
\bibitem [{\citenamefont {Iwai}\ and\ \citenamefont
  {Okamoto}(2006)}]{Iwai2006}%
  \BibitemOpen
  \bibfield  {author} {\bibinfo {author} {\bibfnamefont {S.}~\bibnamefont
  {Iwai}}\ and\ \bibinfo {author} {\bibfnamefont {H.}~\bibnamefont {Okamoto}},\
  }\bibfield  {title} {\bibinfo {title} {Ultrafast {P}hase {C}ontrol in
  {O}ne-{D}imensional {C}orrelated {E}lectron {S}ystems},\ }\href
  {https://doi.org/10.1143/JPSJ.75.011007} {\bibfield  {journal} {\bibinfo
  {journal} {J. Phys. Soc. Jpn.}\ }\textbf {\bibinfo {volume} {75}},\ \bibinfo
  {pages} {011007} (\bibinfo {year} {2006})}\BibitemShut {NoStop}%
\bibitem [{\citenamefont {Giannetti}\ \emph {et~al.}(2016)\citenamefont
  {Giannetti}, \citenamefont {Capone}, \citenamefont {Fausti}, \citenamefont
  {Fabrizio}, \citenamefont {Parmigiani},\ and\ \citenamefont
  {Mihailovic}}]{Giannetti2016}%
  \BibitemOpen
  \bibfield  {author} {\bibinfo {author} {\bibfnamefont {C.}~\bibnamefont
  {Giannetti}}, \bibinfo {author} {\bibfnamefont {M.}~\bibnamefont {Capone}},
  \bibinfo {author} {\bibfnamefont {D.}~\bibnamefont {Fausti}}, \bibinfo
  {author} {\bibfnamefont {M.}~\bibnamefont {Fabrizio}}, \bibinfo {author}
  {\bibfnamefont {F.}~\bibnamefont {Parmigiani}},\ and\ \bibinfo {author}
  {\bibfnamefont {D.}~\bibnamefont {Mihailovic}},\ }\bibfield  {title}
  {\bibinfo {title} {Ultrafast optical spectroscopy of strongly correlated
  materials and high-temperature superconductors: a non-equilibrium approach},\
  }\href {https://doi.org/10.1080/00018732.2016.1194044} {\bibfield  {journal}
  {\bibinfo  {journal} {Adv. Phys.}\ }\textbf {\bibinfo {volume} {65}},\
  \bibinfo {pages} {58} (\bibinfo {year} {2016})}\BibitemShut {NoStop}%
\bibitem [{\citenamefont {Koshihara}\ \emph {et~al.}(2022)\citenamefont
  {Koshihara}, \citenamefont {Ishikawa}, \citenamefont {Okimoto}, \citenamefont
  {Onda}, \citenamefont {Fukaya}, \citenamefont {Hada}, \citenamefont
  {Hayashi}, \citenamefont {Ishihara},\ and\ \citenamefont
  {Luty}}]{Koshihara2022}%
  \BibitemOpen
  \bibfield  {author} {\bibinfo {author} {\bibfnamefont {S.}~\bibnamefont
  {Koshihara}}, \bibinfo {author} {\bibfnamefont {T.}~\bibnamefont {Ishikawa}},
  \bibinfo {author} {\bibfnamefont {Y.}~\bibnamefont {Okimoto}}, \bibinfo
  {author} {\bibfnamefont {K.}~\bibnamefont {Onda}}, \bibinfo {author}
  {\bibfnamefont {R.}~\bibnamefont {Fukaya}}, \bibinfo {author} {\bibfnamefont
  {M.}~\bibnamefont {Hada}}, \bibinfo {author} {\bibfnamefont {Y.}~\bibnamefont
  {Hayashi}}, \bibinfo {author} {\bibfnamefont {S.}~\bibnamefont {Ishihara}},\
  and\ \bibinfo {author} {\bibfnamefont {T.}~\bibnamefont {Luty}},\ }\bibfield
  {title} {\bibinfo {title} {Challenges for developing photo-induced phase
  transition ({PIPT}) systems: {F}rom classical (incoherent) to quantum
  (coherent) control of {PIPT} dynamics},\ }\href
  {https://doi.org/https://doi.org/10.1016/j.physrep.2021.10.003} {\bibfield
  {journal} {\bibinfo  {journal} {Phys. Rep.}\ }\textbf {\bibinfo {volume}
  {942}},\ \bibinfo {pages} {1} (\bibinfo {year} {2022})}\BibitemShut {NoStop}%
\bibitem [{\citenamefont {Ghimire}\ and\ \citenamefont
  {Reis}(2019)}]{Ghimire2019}%
  \BibitemOpen
  \bibfield  {author} {\bibinfo {author} {\bibfnamefont {S.}~\bibnamefont
  {Ghimire}}\ and\ \bibinfo {author} {\bibfnamefont {D.~A.}\ \bibnamefont
  {Reis}},\ }\bibfield  {title} {\bibinfo {title} {High-harmonic generation
  from solids},\ }\href {https://doi.org/10.1038/s41567-018-0315-5} {\bibfield
  {journal} {\bibinfo  {journal} {Nat. Phys.}\ }\textbf {\bibinfo {volume}
  {15}},\ \bibinfo {pages} {10} (\bibinfo {year} {2019})}\BibitemShut {NoStop}%
\bibitem [{\citenamefont {Shimano}\ and\ \citenamefont
  {Tsuji}(2020)}]{Shimano2020}%
  \BibitemOpen
  \bibfield  {author} {\bibinfo {author} {\bibfnamefont {R.}~\bibnamefont
  {Shimano}}\ and\ \bibinfo {author} {\bibfnamefont {N.}~\bibnamefont
  {Tsuji}},\ }\bibfield  {title} {\bibinfo {title} {Higgs {M}ode in
  {S}uperconductors},\ }\href
  {https://doi.org/https://doi.org/10.1146/annurev-conmatphys-031119-050813}
  {\bibfield  {journal} {\bibinfo  {journal} {Annu. Rev. Condens. Matter
  Phys.}\ }\textbf {\bibinfo {volume} {11}},\ \bibinfo {pages} {103} (\bibinfo
  {year} {2020})}\BibitemShut {NoStop}%
\bibitem [{\citenamefont {Ma}\ \emph {et~al.}(2021)\citenamefont {Ma},
  \citenamefont {Grushin},\ and\ \citenamefont {Burch}}]{Ma2021}%
  \BibitemOpen
  \bibfield  {author} {\bibinfo {author} {\bibfnamefont {Q.}~\bibnamefont
  {Ma}}, \bibinfo {author} {\bibfnamefont {A.~G.}\ \bibnamefont {Grushin}},\
  and\ \bibinfo {author} {\bibfnamefont {K.~S.}\ \bibnamefont {Burch}},\
  }\bibfield  {title} {\bibinfo {title} {Topology and geometry under the
  nonlinear electromagnetic spotlight},\ }\href
  {https://doi.org/10.1038/s41563-021-00992-7} {\bibfield  {journal} {\bibinfo
  {journal} {Nat. Mater.}\ }\textbf {\bibinfo {volume} {20}},\ \bibinfo {pages}
  {1601} (\bibinfo {year} {2021})}\BibitemShut {NoStop}%
\bibitem [{\citenamefont {Bukov}\ \emph {et~al.}(2015)\citenamefont {Bukov},
  \citenamefont {D'Alessio},\ and\ \citenamefont {Polkovnikov}}]{Bukov2015}%
  \BibitemOpen
  \bibfield  {author} {\bibinfo {author} {\bibfnamefont {M.}~\bibnamefont
  {Bukov}}, \bibinfo {author} {\bibfnamefont {L.}~\bibnamefont {D'Alessio}},\
  and\ \bibinfo {author} {\bibfnamefont {A.}~\bibnamefont {Polkovnikov}},\
  }\bibfield  {title} {\bibinfo {title} {Universal high-frequency behavior of
  periodically driven systems: from dynamical stabilization to {F}loquet
  engineering},\ }\href {https://doi.org/10.1080/00018732.2015.1055918}
  {\bibfield  {journal} {\bibinfo  {journal} {Adv. Phys.}\ }\textbf {\bibinfo
  {volume} {64}},\ \bibinfo {pages} {139} (\bibinfo {year} {2015})}\BibitemShut
  {NoStop}%
\bibitem [{\citenamefont {Oka}\ and\ \citenamefont {Kitamura}(2019)}]{Oka2019}%
  \BibitemOpen
  \bibfield  {author} {\bibinfo {author} {\bibfnamefont {T.}~\bibnamefont
  {Oka}}\ and\ \bibinfo {author} {\bibfnamefont {S.}~\bibnamefont {Kitamura}},\
  }\bibfield  {title} {\bibinfo {title} {{F}loquet {E}ngineering of {Q}uantum
  {M}aterials},\ }\href
  {https://doi.org/10.1146/annurev-conmatphys-031218-013423} {\bibfield
  {journal} {\bibinfo  {journal} {Annu. Rev. Condens. Matter Phys.}\ }\textbf
  {\bibinfo {volume} {10}},\ \bibinfo {pages} {387} (\bibinfo {year}
  {2019})}\BibitemShut {NoStop}%
\bibitem [{\citenamefont {Oka}\ and\ \citenamefont {Aoki}(2009)}]{Oka2009}%
  \BibitemOpen
  \bibfield  {author} {\bibinfo {author} {\bibfnamefont {T.}~\bibnamefont
  {Oka}}\ and\ \bibinfo {author} {\bibfnamefont {H.}~\bibnamefont {Aoki}},\
  }\bibfield  {title} {\bibinfo {title} {Photovoltaic {H}all effect in
  graphene},\ }\href {https://doi.org/10.1103/PhysRevB.79.081406} {\bibfield
  {journal} {\bibinfo  {journal} {Phys. Rev. B}\ }\textbf {\bibinfo {volume}
  {79}},\ \bibinfo {pages} {081406} (\bibinfo {year} {2009})}\BibitemShut
  {NoStop}%
\bibitem [{\citenamefont {Kitagawa}\ \emph {et~al.}(2011)\citenamefont
  {Kitagawa}, \citenamefont {Oka}, \citenamefont {Brataas}, \citenamefont
  {Fu},\ and\ \citenamefont {Demler}}]{Kitagawa2011}%
  \BibitemOpen
  \bibfield  {author} {\bibinfo {author} {\bibfnamefont {T.}~\bibnamefont
  {Kitagawa}}, \bibinfo {author} {\bibfnamefont {T.}~\bibnamefont {Oka}},
  \bibinfo {author} {\bibfnamefont {A.}~\bibnamefont {Brataas}}, \bibinfo
  {author} {\bibfnamefont {L.}~\bibnamefont {Fu}},\ and\ \bibinfo {author}
  {\bibfnamefont {E.}~\bibnamefont {Demler}},\ }\bibfield  {title} {\bibinfo
  {title} {Transport properties of nonequilibrium systems under the application
  of light: Photoinduced quantum {H}all insulators without {L}andau levels},\
  }\href {https://doi.org/10.1103/PhysRevB.84.235108} {\bibfield  {journal}
  {\bibinfo  {journal} {Phys. Rev. B}\ }\textbf {\bibinfo {volume} {84}},\
  \bibinfo {pages} {235108} (\bibinfo {year} {2011})}\BibitemShut {NoStop}%
\bibitem [{\citenamefont {Mentink}\ \emph {et~al.}(2015)\citenamefont
  {Mentink}, \citenamefont {Balzer},\ and\ \citenamefont
  {Eckstein}}]{Mentink2015}%
  \BibitemOpen
  \bibfield  {author} {\bibinfo {author} {\bibfnamefont {J.~H.}\ \bibnamefont
  {Mentink}}, \bibinfo {author} {\bibfnamefont {K.}~\bibnamefont {Balzer}},\
  and\ \bibinfo {author} {\bibfnamefont {M.}~\bibnamefont {Eckstein}},\
  }\bibfield  {title} {\bibinfo {title} {Ultrafast and reversible control of
  the exchange interaction in {M}ott insulators},\ }\href
  {https://doi.org/10.1038/ncomms7708} {\bibfield  {journal} {\bibinfo
  {journal} {Nat. Commun.}\ }\textbf {\bibinfo {volume} {6}},\ \bibinfo {pages}
  {6708} (\bibinfo {year} {2015})}\BibitemShut {NoStop}%
\bibitem [{\citenamefont {Sato}\ \emph {et~al.}(2016)\citenamefont {Sato},
  \citenamefont {Takayoshi},\ and\ \citenamefont {Oka}}]{Sato2016}%
  \BibitemOpen
  \bibfield  {author} {\bibinfo {author} {\bibfnamefont {M.}~\bibnamefont
  {Sato}}, \bibinfo {author} {\bibfnamefont {S.}~\bibnamefont {Takayoshi}},\
  and\ \bibinfo {author} {\bibfnamefont {T.}~\bibnamefont {Oka}},\ }\bibfield
  {title} {\bibinfo {title} {Laser-{D}riven {M}ultiferroics and {U}ltrafast
  {S}pin {C}urrent {G}eneration},\ }\href
  {https://doi.org/10.1103/PhysRevLett.117.147202} {\bibfield  {journal}
  {\bibinfo  {journal} {Phys. Rev. Lett.}\ }\textbf {\bibinfo {volume} {117}},\
  \bibinfo {pages} {147202} (\bibinfo {year} {2016})}\BibitemShut {NoStop}%
\bibitem [{\citenamefont {Kitamura}\ \emph {et~al.}(2017)\citenamefont
  {Kitamura}, \citenamefont {Oka},\ and\ \citenamefont {Aoki}}]{Kitamura2017}%
  \BibitemOpen
  \bibfield  {author} {\bibinfo {author} {\bibfnamefont {S.}~\bibnamefont
  {Kitamura}}, \bibinfo {author} {\bibfnamefont {T.}~\bibnamefont {Oka}},\ and\
  \bibinfo {author} {\bibfnamefont {H.}~\bibnamefont {Aoki}},\ }\bibfield
  {title} {\bibinfo {title} {Probing and controlling spin chirality in {M}ott
  insulators by circularly polarized laser},\ }\href
  {https://doi.org/10.1103/PhysRevB.96.014406} {\bibfield  {journal} {\bibinfo
  {journal} {Phys. Rev. B}\ }\textbf {\bibinfo {volume} {96}},\ \bibinfo
  {pages} {014406} (\bibinfo {year} {2017})}\BibitemShut {NoStop}%
\bibitem [{\citenamefont {Claassen}\ \emph {et~al.}(2017)\citenamefont
  {Claassen}, \citenamefont {Jiang}, \citenamefont {Moritz},\ and\
  \citenamefont {Devereaux}}]{Claassen2017}%
  \BibitemOpen
  \bibfield  {author} {\bibinfo {author} {\bibfnamefont {M.}~\bibnamefont
  {Claassen}}, \bibinfo {author} {\bibfnamefont {H.-C.}\ \bibnamefont {Jiang}},
  \bibinfo {author} {\bibfnamefont {B.}~\bibnamefont {Moritz}},\ and\ \bibinfo
  {author} {\bibfnamefont {T.~P.}\ \bibnamefont {Devereaux}},\ }\bibfield
  {title} {\bibinfo {title} {Dynamical time-reversal symmetry breaking and
  photo-induced chiral spin liquids in frustrated {M}ott insulators},\ }\href
  {https://doi.org/10.1038/s41467-017-00876-y} {\bibfield  {journal} {\bibinfo
  {journal} {Nat. Commun.}\ }\textbf {\bibinfo {volume} {8}},\ \bibinfo {pages}
  {1192} (\bibinfo {year} {2017})}\BibitemShut {NoStop}%
\bibitem [{\citenamefont {Arakawa}\ and\ \citenamefont
  {Yonemitsu}(2021)}]{Arakawa2021}%
  \BibitemOpen
  \bibfield  {author} {\bibinfo {author} {\bibfnamefont {N.}~\bibnamefont
  {Arakawa}}\ and\ \bibinfo {author} {\bibfnamefont {K.}~\bibnamefont
  {Yonemitsu}},\ }\bibfield  {title} {\bibinfo {title} {{F}loquet engineering
  of {M}ott insulators with strong spin-orbit coupling},\ }\href
  {https://doi.org/10.1103/PhysRevB.103.L100408} {\bibfield  {journal}
  {\bibinfo  {journal} {Phys. Rev. B}\ }\textbf {\bibinfo {volume} {103}},\
  \bibinfo {pages} {L100408} (\bibinfo {year} {2021})}\BibitemShut {NoStop}%
\bibitem [{\citenamefont {Kobayashi}\ \emph {et~al.}(2021)\citenamefont
  {Kobayashi}, \citenamefont {Fujiuchi}, \citenamefont {Sugimoto},\ and\
  \citenamefont {Ohta}}]{Kobayashi2021}%
  \BibitemOpen
  \bibfield  {author} {\bibinfo {author} {\bibfnamefont {H.}~\bibnamefont
  {Kobayashi}}, \bibinfo {author} {\bibfnamefont {R.}~\bibnamefont {Fujiuchi}},
  \bibinfo {author} {\bibfnamefont {K.}~\bibnamefont {Sugimoto}},\ and\
  \bibinfo {author} {\bibfnamefont {Y.}~\bibnamefont {Ohta}},\ }\bibfield
  {title} {\bibinfo {title} {Light-induced switching of magnetic order in the
  anisotropic triangular-lattice {H}ubbard model},\ }\href
  {https://doi.org/10.1103/PhysRevB.103.L161106} {\bibfield  {journal}
  {\bibinfo  {journal} {Phys. Rev. B}\ }\textbf {\bibinfo {volume} {103}},\
  \bibinfo {pages} {L161106} (\bibinfo {year} {2021})}\BibitemShut {NoStop}%
\bibitem [{\citenamefont {Shan}\ \emph {et~al.}(2021)\citenamefont {Shan},
  \citenamefont {Ye}, \citenamefont {Chu}, \citenamefont {Lee}, \citenamefont
  {Park}, \citenamefont {Balents},\ and\ \citenamefont {Hsieh}}]{Shan2021}%
  \BibitemOpen
  \bibfield  {author} {\bibinfo {author} {\bibfnamefont {J.-Y.}\ \bibnamefont
  {Shan}}, \bibinfo {author} {\bibfnamefont {M.}~\bibnamefont {Ye}}, \bibinfo
  {author} {\bibfnamefont {H.}~\bibnamefont {Chu}}, \bibinfo {author}
  {\bibfnamefont {S.}~\bibnamefont {Lee}}, \bibinfo {author} {\bibfnamefont
  {J.-G.}\ \bibnamefont {Park}}, \bibinfo {author} {\bibfnamefont
  {L.}~\bibnamefont {Balents}},\ and\ \bibinfo {author} {\bibfnamefont
  {D.}~\bibnamefont {Hsieh}},\ }\bibfield  {title} {\bibinfo {title} {Giant
  modulation of optical nonlinearity by {F}loquet engineering},\ }\href
  {https://doi.org/10.1038/s41586-021-04051-8} {\bibfield  {journal} {\bibinfo
  {journal} {Nature}\ }\textbf {\bibinfo {volume} {600}},\ \bibinfo {pages}
  {235} (\bibinfo {year} {2021})}\BibitemShut {NoStop}%
\bibitem [{\citenamefont {Benito}\ \emph {et~al.}(2014)\citenamefont {Benito},
  \citenamefont {G\'omez-Le\'on}, \citenamefont {Bastidas}, \citenamefont
  {Brandes},\ and\ \citenamefont {Platero}}]{Benito2014}%
  \BibitemOpen
  \bibfield  {author} {\bibinfo {author} {\bibfnamefont {M.}~\bibnamefont
  {Benito}}, \bibinfo {author} {\bibfnamefont {A.}~\bibnamefont
  {G\'omez-Le\'on}}, \bibinfo {author} {\bibfnamefont {V.~M.}\ \bibnamefont
  {Bastidas}}, \bibinfo {author} {\bibfnamefont {T.}~\bibnamefont {Brandes}},\
  and\ \bibinfo {author} {\bibfnamefont {G.}~\bibnamefont {Platero}},\
  }\bibfield  {title} {\bibinfo {title} {{F}loquet engineering of long-range
  $p$-wave superconductivity},\ }\href
  {https://doi.org/10.1103/PhysRevB.90.205127} {\bibfield  {journal} {\bibinfo
  {journal} {Phys. Rev. B}\ }\textbf {\bibinfo {volume} {90}},\ \bibinfo
  {pages} {205127} (\bibinfo {year} {2014})}\BibitemShut {NoStop}%
\bibitem [{\citenamefont {Takasan}\ \emph {et~al.}(2017)\citenamefont
  {Takasan}, \citenamefont {Daido}, \citenamefont {Kawakami},\ and\
  \citenamefont {Yanase}}]{Takasan2017}%
  \BibitemOpen
  \bibfield  {author} {\bibinfo {author} {\bibfnamefont {K.}~\bibnamefont
  {Takasan}}, \bibinfo {author} {\bibfnamefont {A.}~\bibnamefont {Daido}},
  \bibinfo {author} {\bibfnamefont {N.}~\bibnamefont {Kawakami}},\ and\
  \bibinfo {author} {\bibfnamefont {Y.}~\bibnamefont {Yanase}},\ }\bibfield
  {title} {\bibinfo {title} {Laser-induced topological superconductivity in
  cuprate thin films},\ }\href {https://doi.org/10.1103/PhysRevB.95.134508}
  {\bibfield  {journal} {\bibinfo  {journal} {Phys. Rev. B}\ }\textbf {\bibinfo
  {volume} {95}},\ \bibinfo {pages} {134508} (\bibinfo {year}
  {2017})}\BibitemShut {NoStop}%
\bibitem [{\citenamefont {Chono}\ \emph {et~al.}(2020)\citenamefont {Chono},
  \citenamefont {Takasan},\ and\ \citenamefont {Yanase}}]{Chono2020}%
  \BibitemOpen
  \bibfield  {author} {\bibinfo {author} {\bibfnamefont {H.}~\bibnamefont
  {Chono}}, \bibinfo {author} {\bibfnamefont {K.}~\bibnamefont {Takasan}},\
  and\ \bibinfo {author} {\bibfnamefont {Y.}~\bibnamefont {Yanase}},\
  }\bibfield  {title} {\bibinfo {title} {Laser-induced topological $s$-wave
  superconductivity in bilayer transition metal dichalcogenides},\ }\href
  {https://doi.org/10.1103/PhysRevB.102.174508} {\bibfield  {journal} {\bibinfo
   {journal} {Phys. Rev. B}\ }\textbf {\bibinfo {volume} {102}},\ \bibinfo
  {pages} {174508} (\bibinfo {year} {2020})}\BibitemShut {NoStop}%
\bibitem [{\citenamefont {Kitamura}\ and\ \citenamefont
  {Aoki}(2016)}]{Kitamura2016}%
  \BibitemOpen
  \bibfield  {author} {\bibinfo {author} {\bibfnamefont {S.}~\bibnamefont
  {Kitamura}}\ and\ \bibinfo {author} {\bibfnamefont {H.}~\bibnamefont
  {Aoki}},\ }\bibfield  {title} {\bibinfo {title} {$\ensuremath{\eta}$-pairing
  superfluid in periodically-driven fermionic {H}ubbard model with strong
  attraction},\ }\href {https://doi.org/10.1103/PhysRevB.94.174503} {\bibfield
  {journal} {\bibinfo  {journal} {Phys. Rev. B}\ }\textbf {\bibinfo {volume}
  {94}},\ \bibinfo {pages} {174503} (\bibinfo {year} {2016})}\BibitemShut
  {NoStop}%
\bibitem [{\citenamefont {Fujiuchi}\ \emph {et~al.}(2020)\citenamefont
  {Fujiuchi}, \citenamefont {Kaneko}, \citenamefont {Sugimoto}, \citenamefont
  {Yunoki},\ and\ \citenamefont {Ohta}}]{Fujiuchi2020}%
  \BibitemOpen
  \bibfield  {author} {\bibinfo {author} {\bibfnamefont {R.}~\bibnamefont
  {Fujiuchi}}, \bibinfo {author} {\bibfnamefont {T.}~\bibnamefont {Kaneko}},
  \bibinfo {author} {\bibfnamefont {K.}~\bibnamefont {Sugimoto}}, \bibinfo
  {author} {\bibfnamefont {S.}~\bibnamefont {Yunoki}},\ and\ \bibinfo {author}
  {\bibfnamefont {Y.}~\bibnamefont {Ohta}},\ }\bibfield  {title} {\bibinfo
  {title} {Superconductivity and charge density wave under a time-dependent
  periodic field in the one-dimensional attractive {H}ubbard model},\ }\href
  {https://doi.org/10.1103/PhysRevB.101.235122} {\bibfield  {journal} {\bibinfo
   {journal} {Phys. Rev. B}\ }\textbf {\bibinfo {volume} {101}},\ \bibinfo
  {pages} {235122} (\bibinfo {year} {2020})}\BibitemShut {NoStop}%
\bibitem [{\citenamefont {Kennes}\ \emph {et~al.}(2019)\citenamefont {Kennes},
  \citenamefont {Claassen}, \citenamefont {Sentef},\ and\ \citenamefont
  {Karrasch}}]{Kennes2019}%
  \BibitemOpen
  \bibfield  {author} {\bibinfo {author} {\bibfnamefont {D.~M.}\ \bibnamefont
  {Kennes}}, \bibinfo {author} {\bibfnamefont {M.}~\bibnamefont {Claassen}},
  \bibinfo {author} {\bibfnamefont {M.~A.}\ \bibnamefont {Sentef}},\ and\
  \bibinfo {author} {\bibfnamefont {C.}~\bibnamefont {Karrasch}},\ }\bibfield
  {title} {\bibinfo {title} {Light-induced $d$-wave superconductivity through
  {F}loquet-engineered {F}ermi surfaces in cuprates},\ }\href
  {https://doi.org/10.1103/PhysRevB.100.075115} {\bibfield  {journal} {\bibinfo
   {journal} {Phys. Rev. B}\ }\textbf {\bibinfo {volume} {100}},\ \bibinfo
  {pages} {075115} (\bibinfo {year} {2019})}\BibitemShut {NoStop}%
\bibitem [{\citenamefont {Kumar}\ and\ \citenamefont {Lin}(2021)}]{Kumar2021}%
  \BibitemOpen
  \bibfield  {author} {\bibinfo {author} {\bibfnamefont {U.}~\bibnamefont
  {Kumar}}\ and\ \bibinfo {author} {\bibfnamefont {S.-Z.}\ \bibnamefont
  {Lin}},\ }\bibfield  {title} {\bibinfo {title} {Inducing and controlling
  superconductivity in the {H}ubbard honeycomb model using an electromagnetic
  drive},\ }\href {https://doi.org/10.1103/PhysRevB.103.064508} {\bibfield
  {journal} {\bibinfo  {journal} {Phys. Rev. B}\ }\textbf {\bibinfo {volume}
  {103}},\ \bibinfo {pages} {064508} (\bibinfo {year} {2021})}\BibitemShut
  {NoStop}%
\bibitem [{\citenamefont {Kitamura}\ and\ \citenamefont
  {Aoki}(2022)}]{Kitamura2022}%
  \BibitemOpen
  \bibfield  {author} {\bibinfo {author} {\bibfnamefont {S.}~\bibnamefont
  {Kitamura}}\ and\ \bibinfo {author} {\bibfnamefont {H.}~\bibnamefont
  {Aoki}},\ }\bibfield  {title} {\bibinfo {title} {{F}loquet topological
  superconductivity induced by chiral many-body interaction},\ }\href
  {https://doi.org/10.1038/s42005-022-00936-w} {\bibfield  {journal} {\bibinfo
  {journal} {Commun. Phys.}\ }\textbf {\bibinfo {volume} {5}},\ \bibinfo
  {pages} {174} (\bibinfo {year} {2022})}\BibitemShut {NoStop}%
\bibitem [{\citenamefont {Anan}\ \emph {et~al.}(2024)\citenamefont {Anan},
  \citenamefont {Morimoto},\ and\ \citenamefont {Kitamura}}]{Anan2024}%
  \BibitemOpen
  \bibfield  {author} {\bibinfo {author} {\bibfnamefont {T.}~\bibnamefont
  {Anan}}, \bibinfo {author} {\bibfnamefont {T.}~\bibnamefont {Morimoto}},\
  and\ \bibinfo {author} {\bibfnamefont {S.}~\bibnamefont {Kitamura}},\
  }\bibfield  {title} {\bibinfo {title} {Time-dependent {G}utzwiller simulation
  of {F}loquet topological superconductivity},\ }\href
  {https://doi.org/10.1038/s42005-024-01586-w} {\bibfield  {journal} {\bibinfo
  {journal} {Commun. Phys.}\ }\textbf {\bibinfo {volume} {7}},\ \bibinfo
  {pages} {99} (\bibinfo {year} {2024})}\BibitemShut {NoStop}%
\bibitem [{\citenamefont {Fausti}\ \emph {et~al.}(2011)\citenamefont {Fausti},
  \citenamefont {Tobey}, \citenamefont {Dean}, \citenamefont {Kaiser},
  \citenamefont {Dienst}, \citenamefont {Hoffmann}, \citenamefont {Pyon},
  \citenamefont {Takayama}, \citenamefont {Takagi},\ and\ \citenamefont
  {Cavalleri}}]{Fausti2011}%
  \BibitemOpen
  \bibfield  {author} {\bibinfo {author} {\bibfnamefont {D.}~\bibnamefont
  {Fausti}}, \bibinfo {author} {\bibfnamefont {R.~I.}\ \bibnamefont {Tobey}},
  \bibinfo {author} {\bibfnamefont {N.}~\bibnamefont {Dean}}, \bibinfo {author}
  {\bibfnamefont {S.}~\bibnamefont {Kaiser}}, \bibinfo {author} {\bibfnamefont
  {A.}~\bibnamefont {Dienst}}, \bibinfo {author} {\bibfnamefont {M.~C.}\
  \bibnamefont {Hoffmann}}, \bibinfo {author} {\bibfnamefont {S.}~\bibnamefont
  {Pyon}}, \bibinfo {author} {\bibfnamefont {T.}~\bibnamefont {Takayama}},
  \bibinfo {author} {\bibfnamefont {H.}~\bibnamefont {Takagi}},\ and\ \bibinfo
  {author} {\bibfnamefont {A.}~\bibnamefont {Cavalleri}},\ }\bibfield  {title}
  {\bibinfo {title} {Light-{I}nduced {S}uperconductivity in a
  {S}tripe-{O}rdered {C}uprate},\ }\href
  {https://doi.org/10.1126/science.1197294} {\bibfield  {journal} {\bibinfo
  {journal} {Science}\ }\textbf {\bibinfo {volume} {331}},\ \bibinfo {pages}
  {189} (\bibinfo {year} {2011})}\BibitemShut {NoStop}%
\bibitem [{\citenamefont {Mitrano}\ \emph {et~al.}(2016)\citenamefont
  {Mitrano}, \citenamefont {Cantaluppi}, \citenamefont {Nicoletti},
  \citenamefont {Kaiser}, \citenamefont {Perucchi}, \citenamefont {Lupi},
  \citenamefont {Di~Pietro}, \citenamefont {Pontiroli}, \citenamefont
  {Ricc{\`o}}, \citenamefont {Clark}, \citenamefont {Jaksch},\ and\
  \citenamefont {Cavalleri}}]{Mitrano2016}%
  \BibitemOpen
  \bibfield  {author} {\bibinfo {author} {\bibfnamefont {M.}~\bibnamefont
  {Mitrano}}, \bibinfo {author} {\bibfnamefont {A.}~\bibnamefont {Cantaluppi}},
  \bibinfo {author} {\bibfnamefont {D.}~\bibnamefont {Nicoletti}}, \bibinfo
  {author} {\bibfnamefont {S.}~\bibnamefont {Kaiser}}, \bibinfo {author}
  {\bibfnamefont {A.}~\bibnamefont {Perucchi}}, \bibinfo {author}
  {\bibfnamefont {S.}~\bibnamefont {Lupi}}, \bibinfo {author} {\bibfnamefont
  {P.}~\bibnamefont {Di~Pietro}}, \bibinfo {author} {\bibfnamefont
  {D.}~\bibnamefont {Pontiroli}}, \bibinfo {author} {\bibfnamefont
  {M.}~\bibnamefont {Ricc{\`o}}}, \bibinfo {author} {\bibfnamefont {S.~R.}\
  \bibnamefont {Clark}}, \bibinfo {author} {\bibfnamefont {D.}~\bibnamefont
  {Jaksch}},\ and\ \bibinfo {author} {\bibfnamefont {A.}~\bibnamefont
  {Cavalleri}},\ }\bibfield  {title} {\bibinfo {title} {Possible light-induced
  superconductivity in {K}{$_{3}$}{C}{$_{60}$} at high temperature},\ }\href
  {https://doi.org/10.1038/nature16522} {\bibfield  {journal} {\bibinfo
  {journal} {Nature}\ }\textbf {\bibinfo {volume} {530}},\ \bibinfo {pages}
  {461} (\bibinfo {year} {2016})}\BibitemShut {NoStop}%
\bibitem [{\citenamefont {Buzzi}\ \emph {et~al.}(2020)\citenamefont {Buzzi},
  \citenamefont {Nicoletti}, \citenamefont {Fechner}, \citenamefont
  {Tancogne-Dejean}, \citenamefont {Sentef}, \citenamefont {Georges},
  \citenamefont {Biesner}, \citenamefont {Uykur}, \citenamefont {Dressel},
  \citenamefont {Henderson}, \citenamefont {Siegrist}, \citenamefont
  {Schlueter}, \citenamefont {Miyagawa}, \citenamefont {Kanoda}, \citenamefont
  {Nam}, \citenamefont {Ardavan}, \citenamefont {Coulthard}, \citenamefont
  {Tindall}, \citenamefont {Schlawin}, \citenamefont {Jaksch},\ and\
  \citenamefont {Cavalleri}}]{Buzzi2020}%
  \BibitemOpen
  \bibfield  {author} {\bibinfo {author} {\bibfnamefont {M.}~\bibnamefont
  {Buzzi}}, \bibinfo {author} {\bibfnamefont {D.}~\bibnamefont {Nicoletti}},
  \bibinfo {author} {\bibfnamefont {M.}~\bibnamefont {Fechner}}, \bibinfo
  {author} {\bibfnamefont {N.}~\bibnamefont {Tancogne-Dejean}}, \bibinfo
  {author} {\bibfnamefont {M.~A.}\ \bibnamefont {Sentef}}, \bibinfo {author}
  {\bibfnamefont {A.}~\bibnamefont {Georges}}, \bibinfo {author} {\bibfnamefont
  {T.}~\bibnamefont {Biesner}}, \bibinfo {author} {\bibfnamefont
  {E.}~\bibnamefont {Uykur}}, \bibinfo {author} {\bibfnamefont
  {M.}~\bibnamefont {Dressel}}, \bibinfo {author} {\bibfnamefont
  {A.}~\bibnamefont {Henderson}}, \bibinfo {author} {\bibfnamefont
  {T.}~\bibnamefont {Siegrist}}, \bibinfo {author} {\bibfnamefont {J.~A.}\
  \bibnamefont {Schlueter}}, \bibinfo {author} {\bibfnamefont {K.}~\bibnamefont
  {Miyagawa}}, \bibinfo {author} {\bibfnamefont {K.}~\bibnamefont {Kanoda}},
  \bibinfo {author} {\bibfnamefont {M.-S.}\ \bibnamefont {Nam}}, \bibinfo
  {author} {\bibfnamefont {A.}~\bibnamefont {Ardavan}}, \bibinfo {author}
  {\bibfnamefont {J.}~\bibnamefont {Coulthard}}, \bibinfo {author}
  {\bibfnamefont {J.}~\bibnamefont {Tindall}}, \bibinfo {author} {\bibfnamefont
  {F.}~\bibnamefont {Schlawin}}, \bibinfo {author} {\bibfnamefont
  {D.}~\bibnamefont {Jaksch}},\ and\ \bibinfo {author} {\bibfnamefont
  {A.}~\bibnamefont {Cavalleri}},\ }\bibfield  {title} {\bibinfo {title}
  {Photomolecular {H}igh-{T}emperature {S}uperconductivity},\ }\href
  {https://doi.org/10.1103/PhysRevX.10.031028} {\bibfield  {journal} {\bibinfo
  {journal} {Phys. Rev. X}\ }\textbf {\bibinfo {volume} {10}},\ \bibinfo
  {pages} {031028} (\bibinfo {year} {2020})}\BibitemShut {NoStop}%
\bibitem [{\citenamefont {Sentef}\ \emph {et~al.}(2016)\citenamefont {Sentef},
  \citenamefont {Kemper}, \citenamefont {Georges},\ and\ \citenamefont
  {Kollath}}]{Sentef2016}%
  \BibitemOpen
  \bibfield  {author} {\bibinfo {author} {\bibfnamefont {M.~A.}\ \bibnamefont
  {Sentef}}, \bibinfo {author} {\bibfnamefont {A.~F.}\ \bibnamefont {Kemper}},
  \bibinfo {author} {\bibfnamefont {A.}~\bibnamefont {Georges}},\ and\ \bibinfo
  {author} {\bibfnamefont {C.}~\bibnamefont {Kollath}},\ }\bibfield  {title}
  {\bibinfo {title} {Theory of light-enhanced phonon-mediated
  superconductivity},\ }\href {https://doi.org/10.1103/PhysRevB.93.144506}
  {\bibfield  {journal} {\bibinfo  {journal} {Phys. Rev. B}\ }\textbf {\bibinfo
  {volume} {93}},\ \bibinfo {pages} {144506} (\bibinfo {year}
  {2016})}\BibitemShut {NoStop}%
\bibitem [{\citenamefont {Knap}\ \emph {et~al.}(2016)\citenamefont {Knap},
  \citenamefont {Babadi}, \citenamefont {Refael}, \citenamefont {Martin},\ and\
  \citenamefont {Demler}}]{Knap2016}%
  \BibitemOpen
  \bibfield  {author} {\bibinfo {author} {\bibfnamefont {M.}~\bibnamefont
  {Knap}}, \bibinfo {author} {\bibfnamefont {M.}~\bibnamefont {Babadi}},
  \bibinfo {author} {\bibfnamefont {G.}~\bibnamefont {Refael}}, \bibinfo
  {author} {\bibfnamefont {I.}~\bibnamefont {Martin}},\ and\ \bibinfo {author}
  {\bibfnamefont {E.}~\bibnamefont {Demler}},\ }\bibfield  {title} {\bibinfo
  {title} {Dynamical {C}ooper pairing in nonequilibrium electron-phonon
  systems},\ }\href {https://doi.org/10.1103/PhysRevB.94.214504} {\bibfield
  {journal} {\bibinfo  {journal} {Phys. Rev. B}\ }\textbf {\bibinfo {volume}
  {94}},\ \bibinfo {pages} {214504} (\bibinfo {year} {2016})}\BibitemShut
  {NoStop}%
\bibitem [{\citenamefont {Murakami}\ \emph {et~al.}(2017)\citenamefont
  {Murakami}, \citenamefont {Tsuji}, \citenamefont {Eckstein},\ and\
  \citenamefont {Werner}}]{Murakami2016}%
  \BibitemOpen
  \bibfield  {author} {\bibinfo {author} {\bibfnamefont {Y.}~\bibnamefont
  {Murakami}}, \bibinfo {author} {\bibfnamefont {N.}~\bibnamefont {Tsuji}},
  \bibinfo {author} {\bibfnamefont {M.}~\bibnamefont {Eckstein}},\ and\
  \bibinfo {author} {\bibfnamefont {P.}~\bibnamefont {Werner}},\ }\bibfield
  {title} {\bibinfo {title} {Nonequilibrium steady states and transient
  dynamics of conventional superconductors under phonon driving},\ }\href
  {https://doi.org/10.1103/PhysRevB.96.045125} {\bibfield  {journal} {\bibinfo
  {journal} {Phys. Rev. B}\ }\textbf {\bibinfo {volume} {96}},\ \bibinfo
  {pages} {045125} (\bibinfo {year} {2017})}\BibitemShut {NoStop}%
\bibitem [{\citenamefont {Wang}\ \emph {et~al.}(2018)\citenamefont {Wang},
  \citenamefont {Chen}, \citenamefont {Moritz},\ and\ \citenamefont
  {Devereaux}}]{Wang2018}%
  \BibitemOpen
  \bibfield  {author} {\bibinfo {author} {\bibfnamefont {Y.}~\bibnamefont
  {Wang}}, \bibinfo {author} {\bibfnamefont {C.-C.}\ \bibnamefont {Chen}},
  \bibinfo {author} {\bibfnamefont {B.}~\bibnamefont {Moritz}},\ and\ \bibinfo
  {author} {\bibfnamefont {T.~P.}\ \bibnamefont {Devereaux}},\ }\bibfield
  {title} {\bibinfo {title} {Light-{E}nhanced {S}pin {F}luctuations and
  $d$-{W}ave {S}uperconductivity at a {P}hase {B}oundary},\ }\href
  {https://doi.org/10.1103/PhysRevLett.120.246402} {\bibfield  {journal}
  {\bibinfo  {journal} {Phys. Rev. Lett.}\ }\textbf {\bibinfo {volume} {120}},\
  \bibinfo {pages} {246402} (\bibinfo {year} {2018})}\BibitemShut {NoStop}%
\bibitem [{\citenamefont {Claassen}\ \emph {et~al.}(2019)\citenamefont
  {Claassen}, \citenamefont {Kennes}, \citenamefont {Zingl}, \citenamefont
  {Sentef},\ and\ \citenamefont {Rubio}}]{Claassen2019}%
  \BibitemOpen
  \bibfield  {author} {\bibinfo {author} {\bibfnamefont {M.}~\bibnamefont
  {Claassen}}, \bibinfo {author} {\bibfnamefont {D.~M.}\ \bibnamefont
  {Kennes}}, \bibinfo {author} {\bibfnamefont {M.}~\bibnamefont {Zingl}},
  \bibinfo {author} {\bibfnamefont {M.~A.}\ \bibnamefont {Sentef}},\ and\
  \bibinfo {author} {\bibfnamefont {A.}~\bibnamefont {Rubio}},\ }\bibfield
  {title} {\bibinfo {title} {Universal optical control of chiral
  superconductors and {M}ajorana modes},\ }\href
  {https://doi.org/10.1038/s41567-019-0532-6} {\bibfield  {journal} {\bibinfo
  {journal} {Nat. Phys.}\ }\textbf {\bibinfo {volume} {15}},\ \bibinfo {pages}
  {766} (\bibinfo {year} {2019})}\BibitemShut {NoStop}%
\bibitem [{\citenamefont {Kaneko}\ \emph {et~al.}(2019)\citenamefont {Kaneko},
  \citenamefont {Shirakawa}, \citenamefont {Sorella},\ and\ \citenamefont
  {Yunoki}}]{Kaneko2019}%
  \BibitemOpen
  \bibfield  {author} {\bibinfo {author} {\bibfnamefont {T.}~\bibnamefont
  {Kaneko}}, \bibinfo {author} {\bibfnamefont {T.}~\bibnamefont {Shirakawa}},
  \bibinfo {author} {\bibfnamefont {S.}~\bibnamefont {Sorella}},\ and\ \bibinfo
  {author} {\bibfnamefont {S.}~\bibnamefont {Yunoki}},\ }\bibfield  {title}
  {\bibinfo {title} {Photoinduced $\ensuremath{\eta}$ {P}airing in the
  {H}ubbard {M}odel},\ }\href {https://doi.org/10.1103/PhysRevLett.122.077002}
  {\bibfield  {journal} {\bibinfo  {journal} {Phys. Rev. Lett.}\ }\textbf
  {\bibinfo {volume} {122}},\ \bibinfo {pages} {077002} (\bibinfo {year}
  {2019})}\BibitemShut {NoStop}%
\bibitem [{\citenamefont {Ejima}\ \emph {et~al.}(2020)\citenamefont {Ejima},
  \citenamefont {Kaneko}, \citenamefont {Lange}, \citenamefont {Yunoki},\ and\
  \citenamefont {Fehske}}]{Ejima2020}%
  \BibitemOpen
  \bibfield  {author} {\bibinfo {author} {\bibfnamefont {S.}~\bibnamefont
  {Ejima}}, \bibinfo {author} {\bibfnamefont {T.}~\bibnamefont {Kaneko}},
  \bibinfo {author} {\bibfnamefont {F.}~\bibnamefont {Lange}}, \bibinfo
  {author} {\bibfnamefont {S.}~\bibnamefont {Yunoki}},\ and\ \bibinfo {author}
  {\bibfnamefont {H.}~\bibnamefont {Fehske}},\ }\bibfield  {title} {\bibinfo
  {title} {Photoinduced $\ensuremath{\eta}$-pairing at finite temperatures},\
  }\href {https://doi.org/10.1103/PhysRevResearch.2.032008} {\bibfield
  {journal} {\bibinfo  {journal} {Phys. Rev. Res.}\ }\textbf {\bibinfo {volume}
  {2}},\ \bibinfo {pages} {032008} (\bibinfo {year} {2020})}\BibitemShut
  {NoStop}%
\bibitem [{\citenamefont {Murakami}\ \emph {et~al.}(2022)\citenamefont
  {Murakami}, \citenamefont {Takayoshi}, \citenamefont {Kaneko}, \citenamefont
  {Sun}, \citenamefont {Gole{\v z}}, \citenamefont {Millis},\ and\
  \citenamefont {Werner}}]{Murakami2022}%
  \BibitemOpen
  \bibfield  {author} {\bibinfo {author} {\bibfnamefont {Y.}~\bibnamefont
  {Murakami}}, \bibinfo {author} {\bibfnamefont {S.}~\bibnamefont {Takayoshi}},
  \bibinfo {author} {\bibfnamefont {T.}~\bibnamefont {Kaneko}}, \bibinfo
  {author} {\bibfnamefont {Z.}~\bibnamefont {Sun}}, \bibinfo {author}
  {\bibfnamefont {D.}~\bibnamefont {Gole{\v z}}}, \bibinfo {author}
  {\bibfnamefont {A.~J.}\ \bibnamefont {Millis}},\ and\ \bibinfo {author}
  {\bibfnamefont {P.}~\bibnamefont {Werner}},\ }\bibfield  {title} {\bibinfo
  {title} {Exploring nonequilibrium phases of photo-doped {M}ott insulators
  with generalized {G}ibbs ensembles},\ }\href
  {https://doi.org/10.1038/s42005-021-00799-7} {\bibfield  {journal} {\bibinfo
  {journal} {Commun. Phys.}\ }\textbf {\bibinfo {volume} {5}},\ \bibinfo
  {pages} {23} (\bibinfo {year} {2022})}\BibitemShut {NoStop}%
\bibitem [{\citenamefont {Ueda}\ \emph {et~al.}(2024)\citenamefont {Ueda},
  \citenamefont {Kuroki},\ and\ \citenamefont {Kaneko}}]{Ueda2024}%
  \BibitemOpen
  \bibfield  {author} {\bibinfo {author} {\bibfnamefont {R.}~\bibnamefont
  {Ueda}}, \bibinfo {author} {\bibfnamefont {K.}~\bibnamefont {Kuroki}},\ and\
  \bibinfo {author} {\bibfnamefont {T.}~\bibnamefont {Kaneko}},\ }\bibfield
  {title} {\bibinfo {title} {Photoinduced $\ensuremath{\eta}$-pairing
  correlation in the {H}ubbard ladder},\ }\href
  {https://doi.org/10.1103/PhysRevB.109.075122} {\bibfield  {journal} {\bibinfo
   {journal} {Phys. Rev. B}\ }\textbf {\bibinfo {volume} {109}},\ \bibinfo
  {pages} {075122} (\bibinfo {year} {2024})}\BibitemShut {NoStop}%
\bibitem [{\citenamefont {Gassner}\ \emph {et~al.}(2024)\citenamefont
  {Gassner}, \citenamefont {Weber},\ and\ \citenamefont
  {Claassen}}]{Gassner2024}%
  \BibitemOpen
  \bibfield  {author} {\bibinfo {author} {\bibfnamefont {S.}~\bibnamefont
  {Gassner}}, \bibinfo {author} {\bibfnamefont {C.~S.}\ \bibnamefont {Weber}},\
  and\ \bibinfo {author} {\bibfnamefont {M.}~\bibnamefont {Claassen}},\
  }\bibfield  {title} {\bibinfo {title} {Light-induced switching between
  singlet and triplet superconducting states},\ }\href
  {https://doi.org/10.1038/s41467-024-45949-x} {\bibfield  {journal} {\bibinfo
  {journal} {Nat. Commun.}\ }\textbf {\bibinfo {volume} {15}},\ \bibinfo
  {pages} {1776} (\bibinfo {year} {2024})}\BibitemShut {NoStop}%
\bibitem [{\citenamefont {Cr{\'e}pel}\ and\ \citenamefont
  {Fu}(2021)}]{Crepel2021}%
  \BibitemOpen
  \bibfield  {author} {\bibinfo {author} {\bibfnamefont {V.}~\bibnamefont
  {Cr{\'e}pel}}\ and\ \bibinfo {author} {\bibfnamefont {L.}~\bibnamefont
  {Fu}},\ }\bibfield  {title} {\bibinfo {title} {New mechanism and exact theory
  of superconductivity from strong repulsive interaction},\ }\href
  {https://doi.org/10.1126/sciadv.abh2233} {\bibfield  {journal} {\bibinfo
  {journal} {Sci. Adv.}\ }\textbf {\bibinfo {volume} {7}},\ \bibinfo {pages}
  {eabh2233} (\bibinfo {year} {2021})}\BibitemShut {NoStop}%
\bibitem [{\citenamefont {Cr\'epel}\ \emph {et~al.}(2022)\citenamefont
  {Cr\'epel}, \citenamefont {Cea}, \citenamefont {Fu},\ and\ \citenamefont
  {Guinea}}]{Crepel2022prb}%
  \BibitemOpen
  \bibfield  {author} {\bibinfo {author} {\bibfnamefont {V.}~\bibnamefont
  {Cr\'epel}}, \bibinfo {author} {\bibfnamefont {T.}~\bibnamefont {Cea}},
  \bibinfo {author} {\bibfnamefont {L.}~\bibnamefont {Fu}},\ and\ \bibinfo
  {author} {\bibfnamefont {F.}~\bibnamefont {Guinea}},\ }\bibfield  {title}
  {\bibinfo {title} {Unconventional superconductivity due to interband
  polarization},\ }\href {https://doi.org/10.1103/PhysRevB.105.094506}
  {\bibfield  {journal} {\bibinfo  {journal} {Phys. Rev. B}\ }\textbf {\bibinfo
  {volume} {105}},\ \bibinfo {pages} {094506} (\bibinfo {year}
  {2022})}\BibitemShut {NoStop}%
\bibitem [{\citenamefont {Cr^^c3^^a9pel}\ and\ \citenamefont
  {Fu}(2022)}]{Crepel2022pnas}%
  \BibitemOpen
  \bibfield  {author} {\bibinfo {author} {\bibfnamefont {V.}~\bibnamefont
  {Cr^^c3^^a9pel}}\ and\ \bibinfo {author} {\bibfnamefont {L.}~\bibnamefont
  {Fu}},\ }\bibfield  {title} {\bibinfo {title} {Spin-triplet superconductivity
  from excitonic effect in doped insulators},\ }\href
  {https://doi.org/10.1073/pnas.2117735119} {\bibfield  {journal} {\bibinfo
  {journal} {Proc. Natl. Acad. Sci. USA}\ }\textbf {\bibinfo {volume} {119}},\
  \bibinfo {pages} {e2117735119} (\bibinfo {year} {2022})}\BibitemShut
  {NoStop}%
\bibitem [{\citenamefont {He}\ \emph {et~al.}(2023)\citenamefont {He},
  \citenamefont {Yang}, \citenamefont {Profe}, \citenamefont {Bergholtz},\ and\
  \citenamefont {Kennes}}]{He2023}%
  \BibitemOpen
  \bibfield  {author} {\bibinfo {author} {\bibfnamefont {Y.}~\bibnamefont
  {He}}, \bibinfo {author} {\bibfnamefont {K.}~\bibnamefont {Yang}}, \bibinfo
  {author} {\bibfnamefont {J.~B.}\ \bibnamefont {Profe}}, \bibinfo {author}
  {\bibfnamefont {E.~J.}\ \bibnamefont {Bergholtz}},\ and\ \bibinfo {author}
  {\bibfnamefont {D.~M.}\ \bibnamefont {Kennes}},\ }\bibfield  {title}
  {\bibinfo {title} {Superconductivity of repulsive spinless fermions with
  sublattice potentials},\ }\href
  {https://doi.org/10.1103/PhysRevResearch.5.L012009} {\bibfield  {journal}
  {\bibinfo  {journal} {Phys. Rev. Res.}\ }\textbf {\bibinfo {volume} {5}},\
  \bibinfo {pages} {L012009} (\bibinfo {year} {2023})}\BibitemShut {NoStop}%
\bibitem [{\citenamefont {Cr\'epel}\ \emph {et~al.}(2023)\citenamefont
  {Cr\'epel}, \citenamefont {Guerci}, \citenamefont {Cano}, \citenamefont
  {Pixley},\ and\ \citenamefont {Millis}}]{Crepel2023}%
  \BibitemOpen
  \bibfield  {author} {\bibinfo {author} {\bibfnamefont {V.}~\bibnamefont
  {Cr\'epel}}, \bibinfo {author} {\bibfnamefont {D.}~\bibnamefont {Guerci}},
  \bibinfo {author} {\bibfnamefont {J.}~\bibnamefont {Cano}}, \bibinfo {author}
  {\bibfnamefont {J.~H.}\ \bibnamefont {Pixley}},\ and\ \bibinfo {author}
  {\bibfnamefont {A.}~\bibnamefont {Millis}},\ }\bibfield  {title} {\bibinfo
  {title} {Topological {S}uperconductivity in {D}oped {M}agnetic {M}oir\'e
  {S}emiconductors},\ }\href {https://doi.org/10.1103/PhysRevLett.131.056001}
  {\bibfield  {journal} {\bibinfo  {journal} {Phys. Rev. Lett.}\ }\textbf
  {\bibinfo {volume} {131}},\ \bibinfo {pages} {056001} (\bibinfo {year}
  {2023})}\BibitemShut {NoStop}%
\bibitem [{DPT()}]{DPT}%
  \BibitemOpen
  \href@noop {} {}\bibinfo {note} {In the terminology of
  Ref.~\cite{Crepel2021}, the three types of excitations for $n^{\rm
  B}_{\triangle}=0,1,2$ correspond to the dipole, polaron, and trimer
  excitations.}\BibitemShut {Stop}%
\bibitem [{\citenamefont {Mori}\ \emph {et~al.}(2016)\citenamefont {Mori},
  \citenamefont {Kuwahara},\ and\ \citenamefont {Saito}}]{Mori2016}%
  \BibitemOpen
  \bibfield  {author} {\bibinfo {author} {\bibfnamefont {T.}~\bibnamefont
  {Mori}}, \bibinfo {author} {\bibfnamefont {T.}~\bibnamefont {Kuwahara}},\
  and\ \bibinfo {author} {\bibfnamefont {K.}~\bibnamefont {Saito}},\ }\bibfield
   {title} {\bibinfo {title} {Rigorous {B}ound on {E}nergy {A}bsorption and
  {G}eneric {R}elaxation in {P}eriodically {D}riven {Q}uantum {S}ystems},\
  }\href {https://doi.org/10.1103/PhysRevLett.116.120401} {\bibfield  {journal}
  {\bibinfo  {journal} {Phys. Rev. Lett.}\ }\textbf {\bibinfo {volume} {116}},\
  \bibinfo {pages} {120401} (\bibinfo {year} {2016})}\BibitemShut {NoStop}%
\bibitem [{\citenamefont {Mori}(2023)}]{Mori2023}%
  \BibitemOpen
  \bibfield  {author} {\bibinfo {author} {\bibfnamefont {T.}~\bibnamefont
  {Mori}},\ }\bibfield  {title} {\bibinfo {title} {{F}loquet {S}tates in {O}pen
  {Q}uantum {S}ystems},\ }\href
  {https://doi.org/https://www.annualreviews.org/content/journals/10.1146/annurev-conmatphys-040721-015537}
  {\bibfield  {journal} {\bibinfo  {journal} {Annu. Rev. Condens. Metter.
  Phys.}\ }\textbf {\bibinfo {volume} {14}},\ \bibinfo {pages} {35} (\bibinfo
  {year} {2023})}\BibitemShut {NoStop}%
\bibitem [{\citenamefont {Bukov}\ \emph {et~al.}(2016)\citenamefont {Bukov},
  \citenamefont {Heyl}, \citenamefont {Huse},\ and\ \citenamefont
  {Polkovnikov}}]{Bukov2016}%
  \BibitemOpen
  \bibfield  {author} {\bibinfo {author} {\bibfnamefont {M.}~\bibnamefont
  {Bukov}}, \bibinfo {author} {\bibfnamefont {M.}~\bibnamefont {Heyl}},
  \bibinfo {author} {\bibfnamefont {D.~A.}\ \bibnamefont {Huse}},\ and\
  \bibinfo {author} {\bibfnamefont {A.}~\bibnamefont {Polkovnikov}},\
  }\bibfield  {title} {\bibinfo {title} {Heating and many-body resonances in a
  periodically driven two-band system},\ }\href
  {https://doi.org/10.1103/PhysRevB.93.155132} {\bibfield  {journal} {\bibinfo
  {journal} {Phys. Rev. B}\ }\textbf {\bibinfo {volume} {93}},\ \bibinfo
  {pages} {155132} (\bibinfo {year} {2016})}\BibitemShut {NoStop}%
\bibitem [{\citenamefont {Abanin}\ \emph {et~al.}(2017)\citenamefont {Abanin},
  \citenamefont {De~Roeck}, \citenamefont {Ho},\ and\ \citenamefont
  {Huveneers}}]{Abanin2017}%
  \BibitemOpen
  \bibfield  {author} {\bibinfo {author} {\bibfnamefont {D.~A.}\ \bibnamefont
  {Abanin}}, \bibinfo {author} {\bibfnamefont {W.}~\bibnamefont {De~Roeck}},
  \bibinfo {author} {\bibfnamefont {W.~W.}\ \bibnamefont {Ho}},\ and\ \bibinfo
  {author} {\bibfnamefont {F.}~\bibnamefont {Huveneers}},\ }\bibfield  {title}
  {\bibinfo {title} {Effective {H}amiltonians, prethermalization, and slow
  energy absorption in periodically driven many-body systems},\ }\href
  {https://doi.org/10.1103/PhysRevB.95.014112} {\bibfield  {journal} {\bibinfo
  {journal} {Phys. Rev. B}\ }\textbf {\bibinfo {volume} {95}},\ \bibinfo
  {pages} {014112} (\bibinfo {year} {2017})}\BibitemShut {NoStop}%
\bibitem [{\citenamefont {Kuwahara}\ \emph {et~al.}(2016)\citenamefont
  {Kuwahara}, \citenamefont {Mori},\ and\ \citenamefont
  {Saito}}]{Kuwahara2016}%
  \BibitemOpen
  \bibfield  {author} {\bibinfo {author} {\bibfnamefont {T.}~\bibnamefont
  {Kuwahara}}, \bibinfo {author} {\bibfnamefont {T.}~\bibnamefont {Mori}},\
  and\ \bibinfo {author} {\bibfnamefont {K.}~\bibnamefont {Saito}},\ }\bibfield
   {title} {\bibinfo {title} {{F}loquet^^e2^^80^^93{M}agnus theory and generic
  transient dynamics in periodically driven many-body quantum systems},\ }\href
  {https://doi.org/https://doi.org/10.1016/j.aop.2016.01.012} {\bibfield
  {journal} {\bibinfo  {journal} {Ann. Phys.}\ }\textbf {\bibinfo {volume}
  {367}},\ \bibinfo {pages} {96} (\bibinfo {year} {2016})}\BibitemShut
  {NoStop}%
\bibitem [{\citenamefont {Haldane}(1988)}]{Haldane1988}%
  \BibitemOpen
  \bibfield  {author} {\bibinfo {author} {\bibfnamefont {F.~D.~M.}\
  \bibnamefont {Haldane}},\ }\bibfield  {title} {\bibinfo {title} {Model for a
  {Q}uantum {H}all {E}ffect without {L}andau {L}evels: {C}ondensed-{M}atter
  {R}ealization of the "{P}arity {A}nomaly"},\ }\href
  {https://doi.org/10.1103/PhysRevLett.61.2015} {\bibfield  {journal} {\bibinfo
   {journal} {Phys. Rev. Lett.}\ }\textbf {\bibinfo {volume} {61}},\ \bibinfo
  {pages} {2015} (\bibinfo {year} {1988})}\BibitemShut {NoStop}%
\bibitem [{\citenamefont {Zhou}\ \emph {et~al.}(2023)\citenamefont {Zhou},
  \citenamefont {Egan}, \citenamefont {Kush},\ and\ \citenamefont
  {Franz}}]{Zhou2023}%
  \BibitemOpen
  \bibfield  {author} {\bibinfo {author} {\bibfnamefont {B.~T.}\ \bibnamefont
  {Zhou}}, \bibinfo {author} {\bibfnamefont {S.}~\bibnamefont {Egan}}, \bibinfo
  {author} {\bibfnamefont {D.}~\bibnamefont {Kush}},\ and\ \bibinfo {author}
  {\bibfnamefont {M.}~\bibnamefont {Franz}},\ }\bibfield  {title} {\bibinfo
  {title} {Non-{A}belian topological superconductivity in maximally twisted
  double-layer spin-triplet valley-singlet superconductors},\ }\href
  {https://doi.org/10.1038/s42005-023-01165-5} {\bibfield  {journal} {\bibinfo
  {journal} {Commun. Phys.}\ }\textbf {\bibinfo {volume} {6}},\ \bibinfo
  {pages} {47} (\bibinfo {year} {2023})}\BibitemShut {NoStop}%
\bibitem [{\citenamefont {Eckstein}\ \emph {et~al.}(2017)\citenamefont
  {Eckstein}, \citenamefont {Mentink},\ and\ \citenamefont
  {Werner}}]{Eckstein2017}%
  \BibitemOpen
  \bibfield  {author} {\bibinfo {author} {\bibfnamefont {M.}~\bibnamefont
  {Eckstein}}, \bibinfo {author} {\bibfnamefont {J.~H.}\ \bibnamefont
  {Mentink}},\ and\ \bibinfo {author} {\bibfnamefont {P.}~\bibnamefont
  {Werner}},\ }\href@noop {} {\bibinfo {title} {Designing spin and orbital
  exchange {H}amiltonians with ultrashort electric field transients}} (\bibinfo
  {year} {2017}),\ \Eprint {https://arxiv.org/abs/1703.03269}
  {arXiv:1703.03269} \BibitemShut {NoStop}%
\end{thebibliography}%

\end{document}